\documentclass[journal]{IEEEtran}
\IEEEoverridecommandlockouts
\usepackage{subfigure}
\usepackage{cite}
\usepackage{stfloats}
\usepackage{amsmath,amssymb,amsfonts}
\usepackage{amsthm}
\usepackage{algorithm}               
\usepackage{algorithmic}             
\usepackage{graphicx}
\usepackage{textcomp}
\usepackage{xcolor}
\usepackage{setspace}
\usepackage{subfigure}
\usepackage[caption=false,font=footnotesize]{subfig}
\usepackage{booktabs}
\usepackage{multirow} 
\usepackage{diagbox}
\theoremstyle{plain}
\newtheorem{lemma}{ \textbf{Lemma}}
\newtheorem{theorem}{\textbf{Theorem}}
\newtheorem{coro}{\textbf{Corollary}}

\theoremstyle{definition}
\newtheorem{defi}{\textbf{Definition}}

\theoremstyle{remark}
\newtheorem{rem}{\textbf{Remark}}
\newcommand{\R}[1]{\textcolor{black}{#1}}

\begin{document}
\bstctlcite{IEEEexample:BSTcontrol}

\title{A Tractable Approach for Queueing Analysis on Buffer-Aware Scheduling
	
	\thanks{This work is supported in part by the NSFC/RGC Joint Research Scheme under Grant No. 62261160390/N\_HKUST656/22, and in part by the National Natural Science Foundation of China under Grant No. 62471276.}
	
\thanks{Lintao Li and Wei Chen are with the Department of Electronic Engineering, Tsinghua University, Beijing 100084, China. They are also with the State Key Laboratory of Space Network and Communications, Beijing 100084, China, and the Beijing National Research Center for Information Science and Technology, Beijing 100084, China (email: llt20@mails.tsinghua.edu.cn; wchen@tsinghua.edu.cn).}
	
}

\author{\IEEEauthorblockN{Lintao Li, \emph{Graduate Student Member, IEEE},  and Wei Chen, \emph{Senior Member, IEEE}}
}

\maketitle

\begin{abstract}
Low-latency communication has recently attracted considerable attention owing to its potential of enabling delay-sensitive services in next-generation industrial cyber-physical systems. To achieve target average or maximum delay given random arrivals and time-varying channels, buffer-aware scheduling is expected to play a vital role. Evaluating and optimizing buffer-aware scheduling relies on its queueing analysis, while existing tools are not sufficiently tractable. Particularly, Markov chain and Monte-Carlo based approaches are computationally intensive, while large deviation theory (LDT) and extreme value theory (EVT) fail in providing satisfactory accuracy in the small-queue-length (SQL) regime. To tackle these challenges, a tractable yet accurate queueing analysis is presented by judiciously bridging Markovian analysis for the computationally manageable SQL regime and LDT/EVT for large-queue-length (LQL) regime where approximation error diminishes asymptotically. Specifically, we leverage censored Markov chain augmentation to approximate the original one in the SQL regime, while a piecewise approach is conceived to apply LDT/EVT across various queue-length intervals with different scheduling parameters. Furthermore, we derive closed-form bounds on approximation errors, validating the rigor and accuracy of our approach. As a case study, the approach is applied to analytically analyze a Lyapunov-drift-based cross-layer scheduling for wireless transmissions. Numerical results demonstrate its potential in balancing accuracy and complexity.
\end{abstract}

	\begin{IEEEkeywords}
	Low-latency communication, buffer-aware scheduling, queueing analysis, computational probability, censored Markov chain, large deviation theory, extreme value theory, piecewise approximation, tractable approach.
\end{IEEEkeywords}

\section{Introduction}

Latency is a critical metric in wireless communications research. From ultra-reliable and low-latency communications (URLLC) in fifth-generation (5G) networks to hyper-reliable and low-latency communications (HRLLC) in sixth-generation (6G) networks \cite{ITU}, increasingly stringent latency requirements pose significant challenges to the design of wireless systems. Beyond URLLC and HRLLC, latency critically affects the stability of wireless networked control systems (WNCSs) and user experience in immersive communications \cite{ITU,petar2022}. Since achieving bounded latency is inherently challenging \cite{li2023}, it is essential to account for the tail distribution of latency and queue length to ensure the quality of service (QoS) \cite{urllc2}. Consequently, effective scheduling schemes and efficient characterization of the distribution of latency and queue length are fundamental for enabling subsequent design tasks, such as control and computation, in the future wireless systems.

\subsection{Motivation and Limitations of Existing Methods}

Cross-layer scheduling is widely recognized as a promising approach to ensuring the QoS performance of wireless systems \cite{she2021}. This scheduling policy determines the transmission rate by considering the states of the physical, link, and higher layers, often incorporating buffer states, such as queue length, into the decision-making process. Consequently, it falls under the category of buffer-aware scheduling. Existing studies have explored various cross-layer scheduling policies over fading channels, which can be broadly categorized into two types based on their optimality. The first type leverages Markov decision processes (MDPs) to achieve optimal scheduling policies \cite{puterman}. MDP-based methods can be implemented via numerical approaches, such as value and policy iteration algorithms, or through linear programming (LP) methods \cite{zhao2020,Li2024iotj}. However, the high computational complexity of MDP-based methods limits their applicability in large-scale systems. The second type focuses on sub-optimal scheduling policies, which reduce computational complexity while approaching optimal performance asymptotically. Representative methods include Lyapunov optimization \cite{lya}, Whittle’s index-based optimization \cite{whittle}, and deep learning-based scheduling techniques \cite{she2021}.

After determining the cross-layer scheduling policy, analyzing the tail distribution of queue length and latency becomes crucial for evaluating and controlling the occurrence of extreme events. Due to the difficulties of analyzing the tail distribution of latency, existing works often leverage the relationship between latency and queue length, shifting the focus to the tail distribution of queue length~\cite{li2023}. For MDP-based methods, the tail distribution can be derived by solving the stationary probabilities through the system's state transition matrix (STM). However, the high dimension of the STM in practical systems makes this process computationally prohibitive. Additionally, with the rapid advancement of hardware, buffer sizes have grown significantly, approaching infinity in practical scenarios. This further complicates the calculation of stationary probabilities using infinite-dimensional STMs. Similar challenges arise for sub-optimal scheduling policies, where deriving stationary probabilities remains infeasible or time-consuming. Monte Carlo (MC) simulations are commonly employed for approximating stationary probabilities, but the stringent requirements on queue-length violation probability (QVP) or latency violation probability (LVP) in HRLLC (e.g., less than $10^{-9}$) demand an extensive number of simulations, which is also computationally expensive. While queueing theory offers precise characterizations of tail distributions~\cite{queue}, its application is often restricted to specific scenarios, limiting its practicality. Additionally, existing computational probability methods lack relevant conclusions for this topic~\cite{cp}, primarily because buffer-aware scheduling policies lack the structured properties needed to effectively utilize transform-domain techniques or derive characteristic roots.

\begin{table}[t]
	\centering 
	\caption{Acronyms used in the paper}
		\begin{center}  
			\begin{tabular}{cp{60mm}} 
				\toprule
				\textbf{Abbreviation} & \textbf{Definition} \\ \midrule
				SQL & small-queue-length  \\
				LQL & large-queue-length \\
				LDT & large deviation theory \\
				EVT & extreme value theory \\
				STM & state transition matrix \\
				QVP & queue-length-violation probability\\
				EC & effective capacity \\
				DTMC & discrete-time Markov chain \\
				MC & Monte-Carlo\\
				GEV & generalized extreme value \\
				GPD & generalized Pareto distribution \\
				QSI & queue state information \\
				CSI & channel state information \\
				\bottomrule
			\end{tabular}  
	\end{center} 
\end{table}

To efficiently characterize QVP and LVP, large deviation theory (LDT) \cite{ldt} and extreme value theory (EVT) \cite{haan2006} are extensively employed in existing research. Both LDT and EVT offer simplified analytical approximations of QVP and LVP \cite{liutwc2019,Li2024tit}. Notable LDT-based approaches include effective bandwidth \cite{eb} and effective capacity (EC) \cite{ec}. However, these methods have limited applicability. For instance, LDT and EVT achieve high accuracy in the large-queue-length (LQL) regime but lack sufficient consideration in the small-queue-length (SQL) regime \cite{urllc2}. Moreover, LDT is not universally applicable to all cross-layer scheduling policies \cite{li2023}. Even when LDT is applicable, practical challenges remain. Specifically, cross-layer scheduling, which belongs to the buffer-aware scheduling, couples rate selection with queue length of the buffer. Utilizing LDT requires the expectation of rate-related variables, which is analytically infeasible without precise queue-length distributions. This issue commonly arises in systems with variable service rates tied to buffer states, such as routers employing active queue management methods like random early detection (RED) \cite{net}. Therefore, directly employing LDT to approximate QVP and LVP in wireless systems with buffer-aware scheduling is often impractical. \emph{In summary, all above results indicate that there are difficulties on carrying out queueing analysis for buffer-aware scheduling.}

\begin{table}[t]
	\centering
	\small 
	\caption{Main Notation}
	\begin{center} 
		\begin{tabular}{|c|c|} 
			\hline  
			\multicolumn{1}{|c|}{\textbf{Symbol}} & \multicolumn{1}{c|}{\textbf{Definition}} \\ \hline  
			$|h[n]|^2$ & channel power gain  \\ \hline  
			$T$ & coherence time\\ \hline
			$B$ & bandwidth \\ \hline
			$f(\cdot)$ & distribution of	$|h[n]|^2$ \\ \hline
			$\gamma[n]$ & normalized transmit power \\ \hline
			$a[n]$ & the number of arrival packets \\ \hline
			$A$ & size of each packet \\ \hline
			$p_k$ & $\Pr\{a[n]=k\}$ \\ \hline
			$\lambda$ & average arrival rate \\ \hline
			$q[n]$ & queue length \\ \hline
			$\kappa_q$ & the granularity of transmission control \\ \hline
			$\varepsilon(q_{\rm th})$& queue-length-violation probability \\ \hline 
			$\Xi_q^{h}$ & cross-layer scheduling policy \\ \hline $\zeta_k$ & the queue-length threshold of $\Xi_q^{h}$ \\ \hline 
			$\Lambda_k(\cdot)$ & the scheduling policy on $\left[\zeta_k,\zeta_{k+1}\right)$ \\ \hline
			$\boldsymbol{Q}$ & one-step transition probability matrix \\ \hline 
			
			$\boldsymbol{\pi}$ & stationary probability vector of queue length \\ \hline 
			$\zeta$, $\mu$, $\sigma$ & parameters of GEV distribution \\ \hline
			$\tilde{\zeta}$, $\tilde{\sigma}$, $d$ & parameters of GPD \\ \hline
			$\alpha$ & truncation threshold \\ \hline
			$\mathcal{A}$ & $\left\{0,1,\cdots, \alpha\right\}$ \\ \hline
			$\mathcal{B}$ & $\left\{\alpha+1,\alpha+2,\cdots\right\}$ \\ \hline
			$\boldsymbol{Q}^{\mathcal{A}}$ & one-step transition probability matrix of \\
			& censored Markov chain \\ \hline 
			$\overline{\boldsymbol{Q}}^{\mathcal{A}}$ & one-step transition probability matrix of \\
			& last-column augmentation chain \\ \hline 
			$\underline{\boldsymbol{Q}}^{\mathcal{A}}$ & one-step transition probability matrix of \\
			& first-column augmentation chain \\ \hline 
			$\Theta(\alpha)$ & minimum $l_1$ norm of errors given $\alpha$\\ \hline
			$\theta$ & quality-of-service exponent \\ \hline
			$\Psi_{\alpha}^{l}$, $\Psi_{\alpha}^{u}$ & the approximation of $\Pr\left\{q[n]\geq \alpha\right\}$ \\
			&based on $\overline{\boldsymbol{Q}}^{\mathcal{A}}$ and $\underline{\boldsymbol{Q}}^{\mathcal{A}}$, respectively \\ \hline
			$\vartheta(\alpha)$ & accumulated error of the decay rate \\ \hline 
			$\delta$ & scheduling granularity of the queue length \\ \hline 
			$V$ & hyper-parameter of Lyapunov drift \\ \hline 
		\end{tabular}    
	\end{center} 
\end{table}

\subsection{Our Approach and Contributions}

To address these challenges, we propose a unified framework and tractable approach for carrying out queueing analysis in wireless systems adopting buffer-aware scheduling. Specifically, we divide the queue length into two regimes: the SQL regime and the LQL regime. For the SQL regime, we approximate QVP using a truncated STM, employing techniques such as first-column augmentation, last-column augmentation, and censored Markov chain-based augmentation. Among these, censored Markov chain-based augmentation was proved to have superior asymptotic accuracy in the existing work. Since direct derivation of the censored Markov chain is infeasible, we leverage its stochastic upper and lower bounds, which also serve as bounds for the actual QVP asymptotically under large truncation thresholds. For the LQL regime, we introduce a piecewise analysis method that segments the LQL into intervals for LDT and EVT-based approximations. We theoretically prove that the accumulated errors between the actual QVP and our proposed approximation are bounded and derive closed-form expressions for these bounds under various SQL and LQL approximation combinations. Numerical and simulation results validate the effectiveness of our approach.

The main contributions are summarized as follows:

\begin{itemize} 
	\item We formulate a unified framework and tractable approach for analytically characterizing the QVP of wireless systems with buffer-aware scheduling policies, encompassing both SQL and LQL regimes. The proposed framework and computational probability approach covers the wireless systems adopting buffer-aware scheduling. 
	\item For SQL regime analysis, we employ censored Markov chain-based augmentation and its stochastic bounds. Additionally, we prove that cross-layer scheduling policies with threshold properties allow direct derivation of stochastic upper bounds of the censored Markov chain using last-column augmentation. 
	\item For the analysis in the LQL regime, we propose a novel piecewise analyzing method. Specifically, we present the details of the piecewise LDT based approximation.
	\item We theoretically prove that the accumulated errors from combining censored Markov chain and LDT or EVT approximations are bounded and provide closed-form expressions for these bounds.
	 \end{itemize}

\subsection{Paper Organization and Notation}

	The paper is organized as follows. Section \ref{SM} presents the system model and an overview of cross-layer scheduling, EC, and EVT. Section III-A discusses SQL regime analysis, while Section III-B addresses LQL regime analysis. In Section IV, we apply the proposed approximation to a Lyapunov-drift-based cross-layer scheduling policy. Section V validates our analysis through numerical and simulation results, and Section VI concludes the paper.
	
		\emph{Notation:} $\mathbb{E}\{x\}$ denotes the expectation of a random variable $x$. $f'(x)$ represents the first-order derivative of $f(x)$. $\mathcal{CN}(0,1)$ denotes the circularly-symmetric complex normal distribution with variance 1. $\mathbb{C}$ denotes the set of complex numbers. $\mathbb{N}+$ and $\mathbb{R}+$ denote the sets of positive natural numbers and positive real numbers, respectively. \R{Lower-case and upper-case boldface letters represent column vectors and matrices, respectively.} \R{$\boldsymbol{1}$ denotes a column vector with all elements equal to one, while $\boldsymbol{E}$ denotes a matrix with all elements equal to one.} $\boldsymbol{I}$ represents the identity matrix. \R{For a matrix $\boldsymbol{X}$ or vector $\boldsymbol{x}$, $\boldsymbol{X}^{\mathsf{T}}$ and $\boldsymbol{x}^{\mathsf{T}}$ denotes the transpose of $\boldsymbol{X}$ and $\boldsymbol{x}$, respectively.}

\section{System Model}\label{SM}

\begin{figure}[t]
	\centerline{\includegraphics[width=9cm]{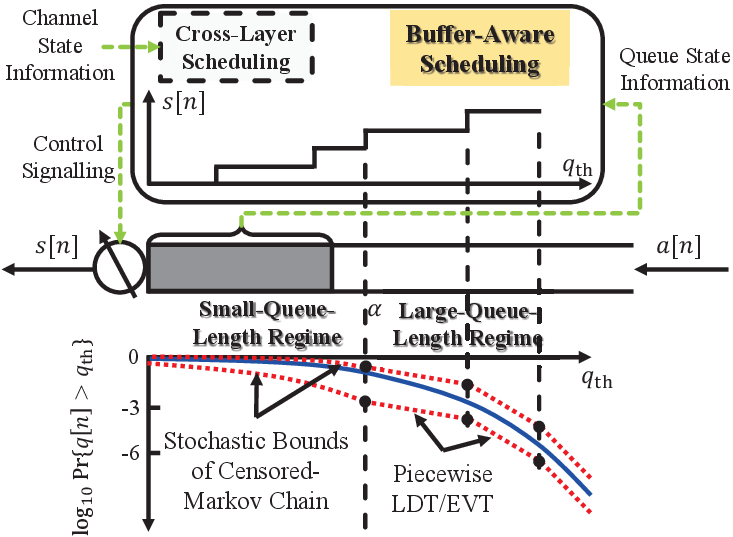}}
	\caption{System model.}
	\label{sysm}
\end{figure}

In this section, we introduce the system model, which considers a classical single-antenna point-to-point communication scenario as illustrated in Fig. \ref{sysm}. The system model encompasses both the physical layer and link layer, as this work investigates cross-layer policies. A general cross-layer policy is then introduced to establish a unified framework for analyzing the tail distribution under arbitrary buffer-aware policies. Additionally, we provide a brief overview of EC and EVT, which are two of widely utilized methods for analyzing the tail distribution of queue length.

\subsection{Physical Layer Model}\label{PLM}

The system operates over a block-fading channel with coherence time $T$. The channel coefficient at time slot $n$ is denoted as $h[n] \in \mathbb{C}$, which remains constant within each time slot but varies independently and identically across time slots. The channel power gain $|h[n]|^2$ follows a distribution with probability density function (PDF) $f(x)$, $x>0$. Given an input signal $\boldsymbol{x}[n] \in \mathbb{C}^N$, the output signal $\boldsymbol{y}[n]\in\mathbb{C}^N$ is expressed as 
\begin{equation}
	\boldsymbol{y}[n]=h[n]\boldsymbol{x}[n]+\boldsymbol{w}[n],
\end{equation}
where $\boldsymbol{w}[n]\sim \mathcal{CN}(0,N_0B\boldsymbol{I})$ represents the additive white Gaussian noise (AWGN) with power spectral density $N_0$, $B$ is the bandwidth, and $N=BT$ is the blocklength. Here we assume that the blocklength is relatively large so that we can neglect the impact of finite blocklength effects \cite{Durisi2016}. The potential revisions due to finite blocklength considerations are remained in the future work.

The normalized transmit power is defined as $\gamma[n] = \frac{P[n]}{N_0 B}$, where $P[n]$ is the transmit power during time slot $n$. \R{Under these assumptions, the channel capacity $C[n]$ (in bits per time slot) for time slot $n$ is given by \cite{shannon}}
\begin{equation}\label{cap}
	C[n]=BT\log_2 \left(1+|h[n]|^2\gamma[n]\right).
\end{equation}

\subsection{Link Layer Model}

 In this subsection, we present the link layer model, focusing on the arrival and service processes within the queuing system at the transmitter's buffer.
 
  The system operates in a time-slotted manner with a slot duration of $T$. An infinite-length buffer is deployed at the transmitter.\footnote{The analysis in this paper is also applicable to finite-length buffers. The infinite-length buffer scenario represents a more general case, as it captures both SQL and LQL regimes.} This buffer allows incoming packets to be stored until they are transmitted. 
  
   For the arrival process, two cases are considered:  
  
  \begin{enumerate}
  	\item \emph{Random Arrival}: In this case, the arrival process is stochastic, with finite mean and variance. Let $a[n]$ denote the number of packets arriving at the buffer at the beginning of time slot $n$, and let $A$ (in bits) represent the size of each packet. The probability of $a[n] = k$ is denoted by $p_k = \Pr\{a[n] = k\}$, where $k = 0, 1, \cdots, a_{\rm max}$ and $a_{\rm max}$ is the maximum number of packets arriving in one slot. The mean arrival rate is given by $\lambda = \sum_{k=0}^{+\infty} k p_k$.
  	\item \emph{Deterministic Arrival}: In this scenario, the number of packets arriving in each time slot is deterministic. Without loss of generality, we let  $a[n] = \lambda$ and the size of each packet is $A$ for all $n$.
  \end{enumerate}

Let $q[n]$ (in packets) denote the queue length at the start of time slot $n$ after the arrival of $a[n]$. We assume that the instantaneous channel state information (CSI) is perfectly known at both the transmitter and receiver, enabling the transmitter to perform adaptive modulation and coding for rate and power control. Under these conditions, the transmitter can send $C[n]$-bit information during time slot $n$. Based on the definition of $C[n]$ in Eq. \eqref{cap}, the number of packets transmitted in time slot $n$, denoted as $s[n]$ (in packets), is given by \cite{wu}
\begin{equation}\label{x3}
	s[n] = \min\left\{q[n], \left\lfloor \frac{C[n]}{\kappa_q A} \right\rfloor\kappa_q , S_{\rm max} \right\},
\end{equation}
where $S_{\rm max}$ is the maximum transmission size constrained by the transmitter's throughput, and $\kappa_q$ represents the granularity of transmission control, which depends on the queue length.\footnote{With advancements in coding and modulation techniques, high-precision control at the bit level is increasingly feasible \cite{Li2024iotj}. Besides, the size of a packet can be large. Therefore, the granularity can be set as relatively continuous.} For simplicity, we assume $\kappa_q = 1$ in the next subsection. As a result, the dynamic equation of the queue length is 
\begin{equation}\label{x4}
	q[n+1]=\max\left\{q[n]-s[n],0\right\}+a[n+1].
\end{equation}

\subsection{Buffer-Aware Scheduling Policy}

In this subsection, we will introduce a cross-layer scheduling policy. This scheduling policy determines the transmission rate given the channel state information (CSI) and queue state information (QSI). Thus, this policy is buffer-aware scheduling policy and takes more elements than the queue state into account.

The goal of this paper is to analyze the tail distribution of the queue length with a buffer-aware scheduling policy. To this end, we define $\pi_m=\lim\limits_{n\to\infty}\Pr\{q[n]=m\}$, $m=0,1,\cdots$, which is the stationary probability of the queue length being equal to $m$. Thus, the QVP with the queue-length threshold $q_{\rm th}$, denoted by $\varepsilon_{q_{\rm th}}$, is given by
\begin{equation}
	\varepsilon\left(q_{\rm th}\right)=\sum_{i=q_{\rm th}+1}^{\infty}\pi_i.
\end{equation}

The cross-layer policy can determine the transmission rate $s[n]$ according to the physical-layer, link-layer, and higher-layer states. In this paper, we take the physical-layer and link-layer states as an instance. Let $\Xi^{h}_{q}$ denote a cross-layer policy, which indicates the amount of transmission packets $s[n]=s$ with the system state $q[n]=q$ and $|h[n]|^2=h$.\footnote{For simplicity, in this paper we consider deterministic cross-layer policies, i.e., $\Pr\left\{\Xi^{h}_{q}=s\right\}\in\{0,1\}$ for all $s$, $h$, and $q$. For stochastic cross-layer policies, the proposed analysis can also be applied.} Note that with a cross-layer policy, the buffer is formulated as a queue with random arrivals and varying service rates.

To better show the influence of the queue length on the cross-layer policy, we use a piecewise function to represent it. Let $\zeta_1,\zeta_2,\cdots,\zeta_K$ denote the thresholds of queue length. These thresholds divide the queue-length space into $K+1$ segments. We let $q\in[\zeta_k,\zeta_{k+1})$, $k\in\{0,1,2,\cdots,K\}$ denote the $k$-th segment, where $\zeta_0=0$ and $\zeta_{K+1}=+\infty$. Based on the above settings, we let $\Lambda_k: h\mapsto s$ with $q\in[\zeta_k,\zeta_{k+1})$, $k\in\{0,1,\cdots,K\}$, denote the scheduling policy in the $k$-th segment. Thus, we have
\begin{equation}\label{pw}
	\begin{aligned}
		\Xi_{q}^{h}=\sum_{k=0}^{K}\mathbb{I}\left\{q\in[\zeta_k,\zeta_{k+1})\right\}\Lambda_k(h),
	\end{aligned}
\end{equation}
where $\mathbb{I}\{\cdot\}$ is the indicator function. Through Eq. \eqref{pw}, we express the cross-layer policy by a piecewise function.

By adopting the cross-layer policy $\{\Xi_q^h\}$, we assume the evolution of the queue length formulates an ergodic discrete-time Markov chain (DTMC). This assumption is satisfied by common policies obtained from Lyapunov optimization \cite{lya} and CMDP \cite{zhao2020}, ensuring the steady state exists for this Markov chain. We let $\mathcal{Q}=\{0,1,2,\cdots\}$ denote the whole space of the queue length, while we let $\boldsymbol{Q}$ denote the one-step transition probability matrix of this DTMC. Moreover, we let $\mathcal{A}=\{0,1,\cdots,\alpha\}$ denote a subspace of $\mathcal{Q}$ with $|\mathcal{A}|$ elements, while $\mathcal{B}=\mathcal{Q}/\mathcal{A}$. $\boldsymbol{Q}$ can be expressed as
\begin{equation}
	\boldsymbol{Q}=\begin{bmatrix}
		\boldsymbol{Q}_{\mathcal{A},\mathcal{A}} & \boldsymbol{Q}_{\mathcal{A},\mathcal{B}}\\
		\boldsymbol{Q}_{\mathcal{B},\mathcal{A}} & \boldsymbol{Q}_{\mathcal{B},\mathcal{B}}
	\end{bmatrix},
\end{equation}
where $\boldsymbol{Q}_{\mathcal{A},\mathcal{A}}$ is a substochastic matrix, which satisfies $\boldsymbol{Q}_{\mathcal{A}, \mathcal{A}} \boldsymbol{1}\leq\boldsymbol{1}$ and $\boldsymbol{Q}_{\mathcal{A}, \mathcal{A}} \boldsymbol{1}\neq \boldsymbol{1}$. The stationary probability vector $\boldsymbol{\pi}$ (column vector) of the DTMC $\left\{q[n]\right\}$ can be solved with $\boldsymbol{Q}$ by
\begin{equation}\label{ssp}
	\boldsymbol{\pi}^{\mathsf{T}}(\boldsymbol{Q}-\boldsymbol{I}+\boldsymbol{E})=\boldsymbol{1}^{\mathsf{T}}.
\end{equation}

\subsection{Classical Methods of Analyzing the Tail Distribution}

In this subsection, we briefly introduce two classical methods for analyzing the tail distribution of queue length: Effective Capacity (EC) and Extreme Value Theory (EVT). 

\subsubsection{Effective Capacity}

EC is a dual concept of Effective Bandwidth (EB), both rooted in LDT. The cornerstone of EC and EB is the Gartner-Ellis theorem \cite{ldt}, which establishes that under certain conditions, the following equation holds
\begin{equation}\label{ttt1}
	\lim_{q_{\rm th} \to +\infty} \frac{\ln\Big(\sup\limits_{n}\big\{\Pr\{q[n] \geq q_{\rm th}\}\big\}\Big)}{q_{\rm th}} = -\theta,
\end{equation}
where $\theta > 0$ is the Quality-of-Service (QoS) exponent \cite{ec}. This exponent represents the decay rate of the queue-length tail distribution. For further details on EC and EB, please refer to \cite{eb,ec}. However, as shown in \cite{li2023}, Eq. \eqref{ttt1} does not always hold for cross-layer policies. Nevertheless, the findings in \cite{li2023} and \cite{Li2024tit} indicate that Eq. \eqref{ttt1} is valid when there is a maximum transmission rate constraint at the transmitter for cross-layer policies.

\subsubsection{Extreme Value Theory}

EVT provides another robust analytical method to characterize the tail distribution of queue length. There are two classical theorems of EVT \cite{urllc2}. We will briefly introduce these two theorems as follows. 

\textbf{(Fisher-Tippett-Gnedenko Theorem)}:  
Given independent samples $X_1, \cdots, X_n$ from the random variable $X$, the cumulative distribution function (CDF) of $M_n = \max\{X_1, \cdots, X_n\}$ \R{is approximated by \cite{urllc2}}:
\begin{equation}\label{evt1}
	\lim_{n\to\infty}\Pr\left\{M_n \leq z\right\} \approx 
	\begin{cases}
		e^{-\left(1+\frac{\xi(z-\mu)}{\sigma}\right)^{-\frac{1}{\xi}}}, & \xi \neq 0, \\
		e^{-e^{-\frac{z-\mu}{\sigma}}}, & \xi = 0,
	\end{cases}
\end{equation}
where $1 + \frac{\xi(z-\mu)}{\sigma} > 0$. The distribution in Eq. \eqref{evt1} is called the generalized extreme value (GEV) distribution.

\textbf{(Pickands-Balkema-De Haan Theorem)}:  
For a random variable $X$ and a sufficiently large threshold $d$, the conditional CDF of the excess value $Y = X - d > 0$ \R{is approximated by \cite{urllc2}}:
\begin{equation}\label{evt2}
	\lim_{d \to \infty} \Pr\{Y < y | X > d\} \approx
	\begin{cases}
		1 - \left(1 + \frac{\tilde{\xi} y}{\tilde{\sigma}}\right)^{-\frac{1}{\tilde{\xi}}}, & \tilde{\xi} > 0, \\
		1 - e^{-\frac{y}{\tilde{\sigma}}}, & \tilde{\xi} = 0,
	\end{cases}
\end{equation}
where $1 + \frac{\tilde{\xi} y}{\tilde{\sigma}} > 0$. If $X$ in Pickands-Balkema-De Haan Theorem follows the same distribution as that of $X$ in the Fisher-Tippett-Gnedenko Theorem, then $\xi = \tilde{\xi}$ and $\tilde{\sigma} = \sigma + \xi(d - \mu)$. The distribution in Eq. \eqref{evt2} is called the generalized Pareto distribution (GPD).\footnote{In this paper, we do not consider the case $\tilde{\xi} < 0$, which corresponds to a bounded distribution where $\exists y > d$, $\Pr\{Y < y | X > d\} = 1$.}

The Pickands-Balkema-De Haan Theorem is typically used to describe the tail distribution of buffer lengths \cite{liutwc2019}. For multiple independent buffers, the Fisher-Tippett-Gnedenko Theorem can characterize the distribution of the maximum queue length among them \cite{liu2019,li2021}. By setting $\tilde{\xi} = 0$ in Eq. \eqref{evt2}, the resulting CDF aligns with Eq. \eqref{ttt1}, demonstrating that EVT is more general than LDT in capturing the tail distribution of queue length.

\section{A Tractable Approach for Characterizing the Tail Distribution of Queue Length}

In this section, we analytically characterize the tail distribution of the queue length under cross-layer scheduling policies by partitioning the queue-length state space into two regimes. The first regime is called SQL regime, which refers to the queue-length interval $[0,\alpha]$. Since the state space of SQL regime remains finite, SQL regime is computationally manageable so that we are permitted to obtain the truncated one-step transition matrix, i.e., $	\boldsymbol{Q}_{\mathcal{A},\mathcal{A}}$, through computational methods. In the SQL regime, we truncate the original chain using a censored Markov chain. To address the challenge of calculating the stationary probability of the censored chain, we employ stochastic bounds to derive upper and lower bounds for the QVP of the censored Markov chain. The second regime is called LQL regime, which refers to the queue-length interval $(\alpha,+\infty)$. LQL regime is amenable to asymptotic analysis as it concentrates on the tail behavior of the queue‐length distribution. In the LQL regime, we propose a piecewise approximation method, which can incorporate with LDT or EVT to capture the decay rate of QVP. By seamlessly combining the approximations in the SQL and LQL regimes, we theoretically quantify the accumulated errors between the actual QVP and approximations and derive the closed-form bounds for these errors.

\subsection{Computationally Manageable SQL regime: Stochastic Bounds of Censored Chain}\label{smallq}

In this subsection, we introduce the first-column augmentation, last-column augmentation, and censored Markov chain-based augmentation with its stochastic upper and lower bounds to approximate the CDF of the original chain in the SQL regime. We first introduce key conclusions of the censored Markov chain and discuss the challenges of deriving its stationary probability with limited information. Then, we explain how to use stochastic bounds to determine the upper and lower bounds of the CDF for the censored chain. Since the first-column and last-column augmentations are closely related to the stochastic bounds of the censored Markov chain, we incorporate these methods during the explanation of the stochastic bounds.

As described in Eq.~\eqref{ssp}, there are two main challenges in obtaining $\boldsymbol{\pi}$. The first challenge lies in constructing $\boldsymbol{Q}$. With cross-layer scheduling policies, the transition probability matrix depends on both CSI and QSI. For infinite buffers, the submatrices $\boldsymbol{Q}_{\mathcal{A},\mathcal{B}}$, $\boldsymbol{Q}_{\mathcal{B},\mathcal{B}}$, and $\boldsymbol{Q}_{\mathcal{B},\mathcal{A}}$ have infinite dimensions, making their formulation computationally prohibitive or even infeasible \cite{cdc}. The second challenge involves computing the inverse of $(\boldsymbol{Q}-\boldsymbol{I}+\boldsymbol{E})$, which has high computational complexity even when $\boldsymbol{Q}$ is known. To address these challenges, we use augmentation methods to truncate $\boldsymbol{Q}$ into a finite-state Markov chain based on limited information $\boldsymbol{Q}_{\mathcal{A},\mathcal{A}}$ \cite{augmen1,stb1}.

Among common augmentation methods, first-column and last-column augmentations are straightforward to implement.\footnote{Ref. \cite{review} provides a comprehensive summary of augmentation methods.} However, censored Markov chain-based augmentation has been shown to provide superior performance in \cite{zhao1996}. A censored Markov chain can approximate countable Markov chains for their limiting behavior \cite{book1}. Following \cite{zhao1996}, we define the censored Markov chain with state space $\mathcal{A}$. Let $\left\{q_{\alpha}[n]\right\}$ represent the stochastic process where the $n$-th transition corresponds to the $n$-th occurrence of the ergodic Markov chain $\left\{q[n]\right\}$ in the set $\mathcal{A}$. In this process, sample paths of $\left\{q_{\alpha}[n]\right\}$ are derived from $\left\{q[n]\right\}$ by omitting all parts in $\mathcal{B}$. The transition probability matrix of $\left\{q_{\alpha}[n]\right\}$ is denoted by $\boldsymbol{Q}^{\mathcal{A}}$.

Two key results about the censored Markov chain are summarized in the following lemmas from~\cite{zhao1996}. The first lemma describes the relationship between stationary probabilities of the original and censored Markov chains.

\begin{lemma} \label{lm1}
	The stationary probability $\left\{\pi_k^{(\alpha)}\right\}$ of the censored chain with truncation threshold $\alpha$ are given by
	\begin{equation}\label{cenchainstprob}
		\pi_{k}^{(\alpha)}=\frac{\pi_k}{\sum_{i\in\mathcal{A}}\pi_i}, \quad  k\in\mathcal{A}.
	\end{equation}
\end{lemma}

Before presenting Lemma~\ref{lm2}, we define the $l_1$ norm of errors between the stationary probabilities of the original infinite chain $\left\{q[n]\right\}$ and an arbitrary finite chain with state space $\mathcal{A}$. Let $\tilde{\pi}_k^{(\alpha)}$ denote the stationary probabilities of the finite chain. The $l_1$ norm of errors is given by \cite{zhao1996}
\begin{equation}
	l_1(\alpha)=\sum_{k=0}^{\alpha}\left|\tilde{\pi}_k^{(\alpha)}-\pi_k\right|+\sum_{k=\alpha+1}^{\infty}\pi_k.
\end{equation}

\begin{lemma} \label{lm2}
	The censored Markov chain $\{q_\alpha[n]\}$ with stationary probabilities $\left\{\pi_k^{(\alpha)}\right\}$ minimizes the $l_1$ norm of error. The minimum $l_1$ norm of error is given by
	\begin{equation}\label{cenchainprop}
		\Theta(\alpha)=\min l_1(\alpha)=2\left(1-\sum_{k=0}^{\alpha}\pi_k\right).
	\end{equation}
\end{lemma}

The above two lemmas illustrate the power of censored Markov chains in approximating the stationary distribution of specific parts of the original queue. However, obtaining the transition probability matrix $\boldsymbol{Q}^{\mathcal{A}}$ remains challenging. This matrix is expressed as \cite{stb1}
\begin{equation}
	\boldsymbol{Q}^{\mathcal{A}}=\boldsymbol{Q}_{\mathcal{A},\mathcal{A}}+\boldsymbol{Q}_{\mathcal{A},\mathcal{B}}\hat{\boldsymbol{Q}}_{\mathcal{B},\mathcal{B}}\boldsymbol{Q}_{\mathcal{B},\mathcal{A}},
\end{equation}
where $\hat{\boldsymbol{Q}}_{\mathcal{B},\mathcal{B}}=\sum_{k=0}^{\infty}{\boldsymbol{Q}}^k_{\mathcal{B},\mathcal{B}}$. Typically, only $\boldsymbol{Q}_{\mathcal{A},\mathcal{A}}$ is available due to its finite dimension and manageable computational cost. Let $\boldsymbol{Q}_{\mathcal{A},\mathcal{A}}^{i,j}$ denote the element in the $i$-th row and $j$-th column of $\boldsymbol{Q}_{\mathcal{A},\mathcal{A}}$, where $i,j \in \{1,2,\cdots,\alpha+1\}$. Under the cross-layer policy $\Xi_q^{h}$, we have
\begin{equation}
	\boldsymbol{Q}_{\mathcal{A},\mathcal{A}}^{i,j}=\sum_{k=0}^{a_{\rm max}}p_k\int_0^{+\infty}  \mathbb{I}\left\{i-\Xi_i^x+k=j\right\} f(x) {\rm d}x.
\end{equation}

As noted in \cite{book1,zhao1996}, directly deriving $\boldsymbol{Q}^{\mathcal{A}}$ is impractical due to limited information about the original Markov chain. A special case arises when $\boldsymbol{Q}$ is an upper Hessenberg matrix, where the censoring operation aligns with the last-column augmentation method, as detailed in \cite{zhao1996}.

Consequently, we cannot easily compute $\pi_k^{(\alpha)}$ with only $\boldsymbol{Q}_{\mathcal{A},\mathcal{A}}$. Instead, we derive bounds for the stationary distribution of the censored Markov chain. Specifically, we calculate the upper and lower bounds for $\sum_{k=i}^{\alpha}\pi_k^{(\alpha)}$, $i \in \mathcal{A}$. To ensure these bounds hold for all $i$, $i \in \mathcal{A}$, we define the strong stochastic order as follows.

\begin{defi} (Strong stochastic order \cite{stb1})  Let $X$ and $Y$ denote two random variables with state space $\{0,1,2,\cdots,N\}$. The probability distribution vector of $X$ is denoted by $\boldsymbol{u}\in \mathbb{R}_+^{1\times (N+1)}$, where $u_i=\Pr\{X=i\}$. Similarly, the probability distribution vector of $Y$ is denoted by $\boldsymbol{v}\in \mathbb{R}_+^{1\times (N+1)}$. The strong stochastic order $X\leq_{\rm st}Y$ holds if and only if
\begin{equation}\label{def1}
	\sum_{i=k}^Nu_i\leq \sum_{i=k}^Nv_i,\quad \forall k\in\{0, 1,2,\cdots,N\}.
\end{equation}
We also write $\boldsymbol{u}\leq_{\rm st}\boldsymbol{v}$ if Eq. \eqref{def1} holds.
\end{defi}

Based on the definition of stochastic orders, we now define the concepts of stochastic monotonicity and stochastic comparability for stochastic matrices, as summarized in Definition \ref{defmono}.

\begin{defi} \label{defmono} (Stochastic monotonicity and comparability \cite{stb1}, \cite{stb2})  Consider a stochastic matrix $\boldsymbol{U}$ of dimension $(N+1) \times (N+1)$. The matrix $\boldsymbol{U}$ satisfies stochastic monotonicity if and only if 
\begin{equation}\label{stmono}
	\boldsymbol{U}_{i,*}\leq_{\rm st}\boldsymbol{U}_{j,*}, \quad \forall i,j \in\{0,1,2,\cdots,N\} \text{ and } i\leq j,
\end{equation}
where $\boldsymbol{U}_{i,*}$ denotes the $i$-row vector of $\boldsymbol{U}$.

For two stochastic matrices $\boldsymbol{U}$ and $\boldsymbol{V}$ of dimension $(N+1)\times(N+1)$, $\boldsymbol{U}$ is stochastically comparable to $\boldsymbol{V}$, denoted as $\boldsymbol{U} \leq_{\rm st} \boldsymbol{V}$, if and only if 
\begin{equation}
	\boldsymbol{U}_{i,*}\leq_{\rm st} \boldsymbol{V}_{i,*}, \quad \forall i\in\{0,1,2,\cdots,N\}.
\end{equation}
\end{defi}

Based on these definitions, we can then introduce Lemma~\ref{lm3}, which provides theoretical guarantees for deriving the stochastic bounds of $\sum_{k=i}^{\alpha}\pi_k^{(\alpha)}$.

\begin{lemma} 
	
	\label{lm3} 
	(Stochastic bounds of stationary probabilities \cite{stb1})
	Let $\{X[n]\}$ and $\{Y[n]\}$ be two DTMCs of order $N$ with transition probability matrices $\boldsymbol{U}$ and $\boldsymbol{V}$, respectively. Suppose their stationary distributions exist and are denoted by $\boldsymbol{u}$ and $\boldsymbol{v}$. The stochastic order $\boldsymbol{u} \leq_{\rm st} \boldsymbol{v}$ holds if and only if the following conditions are satisfied:
	\begin{itemize}
		\item $X[0]\leq_{\rm st}Y[0]$;
		\item either $\boldsymbol{U}$ or $\boldsymbol{V}$ satisfies stochastic monotonicity;
		\item $\boldsymbol{U}\leq_{\rm st}\boldsymbol{V}$.
	\end{itemize}
\end{lemma}

To derive the stochastic upper and lower bounds of $\sum_{k=i}^{\alpha}\pi_k^{(\alpha)}$, we first need to construct two transition probability matrices, $\underline{\boldsymbol{Q}}^{\mathcal{A}}$ and $\overline{\boldsymbol{Q}}^{\mathcal{A}}$, such that
\begin{equation}\label{sto}
	\underline{\boldsymbol{Q}}^{\mathcal{A}}\leq_{\rm st} \boldsymbol{Q}^{\mathcal{A}} \leq_{\rm st} \overline{\boldsymbol{Q}}^{\mathcal{A}}.
\end{equation}
As stated in \cite{stb1}, the way to obtain $\overline{\boldsymbol{Q}}^{\mathcal{A}}$ can be obtained through last-column augmentation, while $\underline{\boldsymbol{Q}}^{\mathcal{A}}$ can be obtained through first-column augmentation. The procedures for obtaining $\boldsymbol{Q}_{\mathcal{A},\mathcal{A}}$, $\overline{\boldsymbol{Q}}^{\mathcal{A}}$, and $\underline{\boldsymbol{Q}}^{\mathcal{A}}$ are outlined in Algorithm~\ref{alg1}. In Algorithm~\ref{alg1}, let $\boldsymbol{Q}_{\mathcal{A},\mathcal{A}}^{i,j}$ represent the $(i,j)$-th entry of $\boldsymbol{Q}_{\mathcal{A},\mathcal{A}}$, where $i, j \in \{1,2,\cdots,\alpha+1\}$. Moreover, $\boldsymbol{e}_j$ denotes the $j$-th column of the identity matrix of size $(\alpha+1)$.\footnote{Alternative methods to obtain $\underline{\boldsymbol{Q}}^{\mathcal{A}}$ and $\overline{\boldsymbol{Q}}^{\mathcal{A}}$ can be found in Algorithms 5 and 6 in \cite{stb2}.}

	\begin{figure*}[b]
	\normalsize \hrulefill
	\setcounter{equation}{23}
	\begin{equation}\label{thm1p1}
		\begin{aligned}
			\sum_{n=j}^{\infty}Q_{i+1,n}-\sum_{n=j}^{\infty}Q_{i,n}=&\Pr\{q[n+1]\geq j \, |\, q[n]=i+1\}-\Pr \left\{q[n+1]\geq j \, |\, q[n]=i\right\}\\
			=&\sum_{k=0}^{a_{\rm max}}p_k\int_0^{+\infty}\Big(\mathbb{I}\left\{\Xi_{i+1}^x\leq i+1+k-j\right\}-\mathbb{I}\left\{\Xi_i^x\leq i+k-j\right\}\Big)f(x){\rm d}x,
		\end{aligned}
	\end{equation}
\end{figure*}

Since $\boldsymbol{Q}^{\mathcal{A}}$ may not satisfy stochastic monotonicity, we adjust $\underline{\boldsymbol{Q}}^{\mathcal{A}}$ and $\overline{\boldsymbol{Q}}^{\mathcal{A}}$ to ensure they do so that we can use Lemma~\ref{lm3} to derive the stochastic bounds. The adjusted matrices, which satisfy Eq.~\eqref{sto} and are stochastically monotonic, are denoted as $\underline{\boldsymbol{S}}^{\mathcal{A}}$ and $\overline{\boldsymbol{S}}^{\mathcal{A}}$, which are obtained from $\underline{\boldsymbol{Q}}^{\mathcal{A}}$ and $\overline{\boldsymbol{Q}}^{\mathcal{A}}$, respectively. Due to space limitations, we refer readers to Algorithms 3 and 4 in \cite{stb2} for detailed implementation. After obtaining $\overline{\boldsymbol{S}}^{\mathcal{A}}$, we can solve its corresponding stationary probability from the following steady-state equation:
	\setcounter{equation}{20}
\begin{equation}\label{upperpi}
	\boldsymbol{\pi}_{u}^{\mathsf{T}} =\boldsymbol{1}^{\mathsf{T}}\left(\overline{\boldsymbol{S}}^{\mathcal{A}}-\boldsymbol{I}+\boldsymbol{E}\right)^{-1}.
\end{equation}
Based on Eq. \eqref{upperpi}, we can also derive an approximation of $\Pr\left\{q[n]\geq  \alpha\right\}$, which is denoted by $\Psi^u_{\alpha}=1-\sum_{i=0}^{\alpha-1}\pi_{u,i}$. Similarly, we can obtain the stationary probability of $\overline{\boldsymbol{S}}^{\mathcal{A}}$ by
\begin{equation}\label{lowerpi}
	\boldsymbol{\pi}_{l}^{\mathsf{T}} =\boldsymbol{1}^{\mathsf{T}}\left(\underline{\boldsymbol{S}}^{\mathcal{A}}-\boldsymbol{I}+\boldsymbol{E}\right)^{-1}.
\end{equation}
We then obtain another approximation of $\Pr\left\{q[n]\geq  \alpha\right\}$, which is denoted by $\Psi^l_{\alpha}=1-\sum_{i=0}^{\alpha-1}\pi_{l,i}$.

\renewcommand{\algorithmicrequire}{\textbf{Input:}}  
\renewcommand{\algorithmicensure}{\textbf{Output:}}  
\begin{algorithm}[t]
	{
		\caption{Construct $\underline{\boldsymbol{Q}}^{\mathcal{A}}$ and $\overline{\boldsymbol{Q}}^{\mathcal{A}}$}
		\begin{algorithmic}[1] \label{alg1}
			\REQUIRE
			cross-layer policy $\left\{\Xi_q^h\right\}$ and truncation threshold $\alpha$
			\ENSURE
			$\underline{\boldsymbol{Q}}^{\mathcal{A}}$ and $\overline{\boldsymbol{Q}}^{\mathcal{A}}$
			\FORALL{$i,j\in\{1,2,\cdots,\alpha+1\}$}
			\STATE 	$\boldsymbol{Q}_{\mathcal{A},\mathcal{A}}^{i,j}\leftarrow\sum_{k=0}^{a_{\rm max}}p_k\int_0^{+\infty}  \mathbb{I}\left\{i-\Xi_i^x+k=j\right\} f(x) {\rm d}x $
			\ENDFOR
			\STATE $\Delta_{\mathcal{A}}\leftarrow\boldsymbol{1}-\boldsymbol{Q}_{\mathcal{A},\mathcal{A}}\boldsymbol{1}$
			\STATE $\overline{\boldsymbol{Q}}^{\mathcal{A}}\leftarrow  \boldsymbol{Q}_{\mathcal{A},\mathcal{A}}+\Delta_{\mathcal{A}}\boldsymbol{e}_{\alpha+1}^{\mathsf{T}}$ 
			\STATE 
			$\underline{\boldsymbol{Q}}^{\mathcal{A}}\leftarrow \boldsymbol{Q}_{\mathcal{A},\mathcal{A}}+ \Delta_{\mathcal{A}}\boldsymbol{e}_{1}^{\mathsf{T}}$ 
			\RETURN 
			$\underline{\boldsymbol{Q}}^{\mathcal{A}}$ and $\overline{\boldsymbol{Q}}^{\mathcal{A}}$
	\end{algorithmic}}
\end{algorithm}

In \cite{zhao2020}, the optimal cross-layer scheduling policy for minimizing the average latency was shown to exhibit a threshold property. Similar property was proved in many works. This property indicates that the optimal cross-layer scheduling policy $\Xi_q^{h}$ is monotonically increasing with $q$, which means $\Xi_{q+1}^h-\Xi_q^h\geq 0$. Additionally, it also shows that $\Xi_{q+1}^h-\Xi_q^h\leq 1$. Leveraging these properties, we prove in Theorem~\ref{thm1} that when a cross-layer scheduling policy following the threshold property is employed, $\overline{\boldsymbol{Q}}^{\mathcal{A}}$ obtained from Algorithm~\ref{alg1} is stochastically monotonic.

\begin{theorem}\label{thm1}
	For a wireless system with $\kappa_q=1$ that adopts a cross-layer scheduling policy following the threshold property, i.e.,
	\setcounter{equation}{22}
	\begin{equation}\label{thm1e1}
		\Xi_{q+1}^h-\Xi_q^{h}\in\{0,1\}, \quad q\in\{0,1,\cdots,\alpha-1\},
	\end{equation}
	the corresponding $\overline{\boldsymbol{Q}}^{\mathcal{A}}$ obtained from Algorithm~\ref{alg1} exhibits stochastic monotonicity, i.e., $\overline{\boldsymbol{Q}}^{\mathcal{A}}=\overline{\boldsymbol{S}}^{\mathcal{A}}$.
\end{theorem}
\begin{IEEEproof}
	To prove this theorem, we start from the analysis of the transition probability of the original chain $\{q[n]\}$ with the threshold property. For an arbitrary $i \in \{0,1,\cdots,\alpha\}$, we have the relationship as shown in Eq.~\eqref{thm1p1} at the bottom of this page, where $j\in\left\{0,1,\cdots\right\}$.

	From Eq. \eqref{thm1p1}, we find that, if $\mathbb{I}\{\Xi_i^x\leq i+k-j\}=1$, $\mathbb{I}\{\Xi_{i+1}^x\leq i+k-j+1\}=1$ holds because under this condition the largest possible value of $\Xi_{i+1}^x$ is $\min\{i+1+k-j,S_{\rm max}\}$ according to Eq. \eqref{thm1e1}. Moreover, if $\mathbb{I}\{\Xi_i^x\leq i+k-j\}=0$, which implies $\Xi_i^x>i+k-j$, $\mathbb{I}\{\Xi_{i+1}^x\leq i+k-j+1\}$ can also be equal to 1 for certain values of $x$. This is because, there is a probability that $\Xi_{i+1}^x=\Xi_i^x=i+1+k-j$, which satisfies $\mathbb{I}\{\Xi_i^x\leq i+k-j\}=0$ and $\mathbb{I}\{\Xi_{i+1}^x\leq i+1+k-j\}=1$. Based on Eq. \eqref{thm1p1}, we conclude that
	\setcounter{equation}{24}
	\begin{equation}\label{thm1p2}
		\begin{aligned}
			\sum_{n=j}^{\alpha}\overline{\boldsymbol{Q}}^{\mathcal{A}}_{i+1,n}-	\sum_{n=j}^{\alpha}\overline{\boldsymbol{Q}}^{\mathcal{A}}_{i,j}=\sum_{n=j}^{\infty}\boldsymbol{Q}_{i+1,n}-\sum_{n=j}^{\infty}\boldsymbol{Q}_{i,n}\geq 0.
		\end{aligned}
	\end{equation} 
	Eq. \eqref{thm1p2} holds because, under the last-column augmentation method, $\overline{\boldsymbol{Q}}^{\mathcal{A}}_{i,\alpha}=\boldsymbol{Q}_{i,\alpha}+\sum_{n=\alpha+1}^{\infty}\boldsymbol{Q}_{i,n}$. This result demonstrates that Eq. \eqref{stmono} is satisfied. According to Definition 2, we have proved that $\overline{\boldsymbol{Q}}^{\mathcal{A}}$ is stochastic monotonic.
\end{IEEEproof}

In Theorem~\ref{thm1}, we establish that for cross-layer scheduling policies adhering to the threshold property, the stochastic upper bound can be directly obtained using the last-column augmentation method. In practical applications, the stochastic upper bound is particularly significant as it facilitates conservative design strategies. Hence, Theorem~\ref{thm1} provides a valuable method for efficiently approximating the stationary distribution in the SQL regime.

\begin{rem}
\emph{Note that the threshold property also holds for cross-layer scheduling policies with uniform quantization, as demonstrated in \cite{Li2024iotj}. For such policies, Theorem~\ref{thm1} remains valid. The key difference lies in the minimum unit of the queue length and service rate, which changes from 1 to the quantization granularity.}
\end{rem}

\begin{rem}
	\emph{The first-column and last-column augmentation methods represent two extreme cases of the queueing system under consideration. Specifically, the first-column augmentation method assumes that the queue is emptied when its length exceeds $\alpha$, whereas the last-column augmentation method drops the portion of packets when $q[n] > \alpha$. These methods are general and can be applied to any censored Markov chain to derive stochastic upper and lower bounds. However, for systems governed by cross-layer scheduling policies with specific structural properties, these methods can be further optimized to achieve bounds with reduced computational complexity. Developing such optimizations is an important direction for future research.}
\end{rem}

\begin{rem}
\emph{We attempted to prove that for a cross-layer policy satisfying the threshold property, the matrix $\underline{\boldsymbol{Q}}^{\mathcal{A}}$ obtained from Algorithm~\ref{alg1} exhibits stochastic monotonicity. However, achieving this result requires additional conditions on the PDF of the arrivals and CSI. Specifically, we derive the following relationship: }
\setcounter{equation}{25}
\begin{equation}
	\sum_{n=j}^{\alpha}\underline{\boldsymbol{Q}}^{\mathcal{A}}_{i+1,n}-\sum_{n=j}^{\alpha}\underline{\boldsymbol{Q}}^{\mathcal{A}}_{i,n}=
	\begin{cases}
		\sum\limits_{n=j}^{\alpha}\boldsymbol{Q}_{i+1,n}-\sum\limits_{n=j}^{\alpha}\boldsymbol{Q}_{i,n}, \, j\neq 0\\
		\sum\limits_{n=0}^{\infty}\boldsymbol{Q}_{i+1,n}-\sum\limits_{n=0}^{\infty}\boldsymbol{Q}_{i,n}, \, j=0.
	\end{cases}
\end{equation}
\emph{For the case $j = 0$, it is evident from Eq.~\eqref{thm1p1} that $\sum\limits_{n=0}^{\infty}\boldsymbol{Q}_{i+1,n}-\sum\limits_{n=0}^{\infty}\boldsymbol{Q}_{i,n}=0 $. For $j \neq 0$, we refer to Eq.~\eqref{stateprob}, which is located at the bottom of the next page. Eq.~\eqref{stateprob} introduces further dependencies on the PDF $f(\cdot)$ and the arrival distribution $p_k$. To ensure $\underline{\boldsymbol{Q}}^{\mathcal{A}}$ exhibits stochastic monotonicity, these distributions must satisfy specific constraints. However, since $f(\cdot)$ and $p_k$ are determined by external factors and cannot be directly controlled, it is challenging to guarantee that Eq.~\eqref{stateprob} remains non-negative in all cases. }
	\end{rem}

\begin{figure*}[b]
	\normalsize \hrulefill
	\setcounter{equation}{26}
	\begin{equation}\label{stateprob}
		\begin{aligned}
			&\sum_{n=j}^{\alpha}\sum_{k=0}^{a_{\rm max}}p_k\int_0^{+\infty}\Big(\mathbb{I}\left\{\Xi_{i+1}^x= i+1+k-n\right\}-\mathbb{I}\left\{\Xi_i^x= i+k-n\right\}\Big)f(x){\rm d}x\\
			=&\sum_{k=0}^{a_{\rm max}}p_k\int_0^{+\infty}\Big(\mathbb{I}\left\{\Xi_{i+1}^x\in\left\{ i+1+k-\alpha,\cdots,i+1+k-j\right\}\right\}-\mathbb{I}\left\{\Xi_i^x\in \left\{ i+k-\alpha,\cdots,i+k-j\right\}\right\}\Big)f(x){\rm d}x
		\end{aligned}
	\end{equation}
\end{figure*}

\subsection{LQL Regime Amenable to Asymptotic Analysis: Piecewise LDT Based Analysis}\label{largeq}

Based on the analysis in the previous subsection, we have obtained an approximation of $\pi_i$ in the SQL regime and the probability that the queue length lies in the LQL regime. In this subsection, we introduce the piecewise analysis for the LQL regime. Before presenting the piecewise analysis, we first prove that for the QVP satisfying LDT or EVT, the accumulate error is bounded by adopting the censored Markov chain-based approximation in the SQL regime. Based on this result, we then propose the tractable approach which combines the censored Markov chain-based approximation in the SQL regime with the piecewise LDT or EVT in the LQL regime.

Before presenting the piecewise analysis, we first prove that the accumulated error between the decay rate of the censored Markov chain and the original chain is bounded in the SQL regime. We define the accumulated error as shown in Definition~\ref{def3}.

\begin{defi} \label{def3} (Accumulated error of the decay rate) We define the accumulated error of the decay rate as the $\ell_1$ norm of the errors between $\ln\Pr\{q[n] \geq k\}$ and $\ln\Pr\{q_{\alpha}[n]\geq k\}$ on $\mathcal{A}$, which is given by 
	\setcounter{equation}{27}
\begin{equation}\label{l1er}
	\vartheta(\alpha)=\sum_{k=0}^{\alpha-1}\left| \ln\left(1-\sum_{i=0}^k\pi_i\right)-\ln\left(1-\sum_{i=0}^{k}\pi_i^{(\alpha)}\right)  \right|. 
\end{equation}
The $\ln(\cdot)$ is used to capture the decay rate of the exponents. For wireless systems following LDT, $\ln\{q[n]\geq q_{\rm th}\}\approx -\theta q_{\rm th}$, which characterizes the decay rate of the tail distribution of the queue length.
\end{defi}

Based on Definition~\ref{def3}, Lemma~\ref{lm1}, and Lemma~\ref{lm2}, we have proved the following conclusion in Theorem~\ref{thm2}.

\begin{theorem}\label{thm2}
    	Assume that $\left\{q[n]\right\}$ satisfies the following equation:
    	\begin{equation}\label{thm2e1}
    		\varepsilon(q_{\rm th})=e^{-\theta q_{\rm th}+b q_{\rm th}^p},
    	\end{equation}
     where $\theta$ is a positive constant, $p\in(-\infty,1)$, and $b$ is a finite constant satisfying $-\theta k+bk^p\leq 0$ for $k=0,1,\cdots$. 
     
     For $p\in(-\infty,0)\cup(0,1)$, we have
     \begin{equation}\label{thm2e2}
     	0\leq \lim_{\alpha\to+\infty}\vartheta(\alpha)\leq \frac{1}{(e^{\theta}-1)(1-e^{-\theta })}.
     \end{equation}
 
 For $p=0$, we have
 \begin{equation}\label{thm2e3}
 	\frac{1}{e^{\theta}-1}\leq \lim_{\alpha\to+\infty}\vartheta(\alpha)\leq \frac{1}{(e^{\theta}-1)(1-e^{-\theta })}.
 \end{equation}
\end{theorem}

\begin{IEEEproof}
See Appendix A.
\end{IEEEproof}

Note that, when $\varepsilon(q_{\rm th})$ satisfies the equation in Eq. \eqref{thm2e1}, LDT holds, as demonstrated by the following limit:
\begin{equation}
	\lim_{q_{\rm th}\to+\infty}\frac{\ln \varepsilon(q_{\rm th})}{q_{\rm th}} = -\theta.
\end{equation}
In the majority of existing literature \cite{ec, li2023, li2021gc, li2021}, LDT is typically used to approximate the tail distribution as $\varepsilon(q_{\rm th}) \approx e^{-\theta q_{\rm th}}$ for large values of $q_{\rm th}$, which corresponds to the specific case where $p = 0$ in Theorem~\ref{thm2}. However, we adopt a more general formulation in which $\varepsilon(q_{\rm th})$ satisfies Eq. \eqref{thm2e1}, thus extending the analysis beyond the case of $p = 0$.

Building on Theorem~\ref{thm2}, we demonstrate that for a queueing process $\{q[n]\}$ that adheres to LDT, applying a censored Markov chain to analyze the stationary distribution over the truncated set of system states, $\mathcal{A}$, results in a bounded error on the decay rate, as defined in Eq. \eqref{l1er}. This provides a theoretical foundation for combining the censored Markov chain and LDT as an approximation of the tail distribution of the queue length.\footnote{As mentioned in Section~\ref{smallq}, we choose to use stochastic bounds to mitigate the complexities involved in directly formulating a censored Markov chain. A crucial avenue for future work is to assess the errors introduced by these stochastic bounds.}

Furthermore, according to Eq. \eqref{thm2e3}, we observe that for $p = 0$, the limit $\lim_{\alpha \to +\infty} \vartheta(\alpha)$ approximates $\frac{1}{e^{\theta} - 1}$ as $\theta$ becomes large. In URLLC scenarios, large values of $\theta$ are common, as they reflect stringent QoS requirements \cite{Li2024tit}. Thus, for the approximation $\varepsilon(q_{\rm th}) \approx e^{-\theta q_{\rm th}}$, we identify an asymptotic value for $\vartheta(\alpha)$ in the regime of large $\theta$. Additionally, as $\theta$ increases, the term $\frac{1}{e^{\theta} - 1}$ diminishes, indicating that $\vartheta(\alpha)$ becomes small in low-latency systems.

\renewcommand{\algorithmicrequire}{\textbf{Input:}}  
\renewcommand{\algorithmicensure}{\textbf{Output:}}  
\begin{algorithm}[t]
	{
		\caption{Approximating the tail distribution of $\{q[n]\}$}
		\begin{algorithmic}[1]\label{alg2}
			\REQUIRE
			cross-layer policy $\left\{\Xi_q^h\right\}$ and truncation threshold $\alpha=\zeta_{K-\ell+1}$
			\ENSURE
			$\varepsilon^{u}(j)$ and $\varepsilon^{l}(j)$, $j\in\{0,1,\cdots\}$
			\STATE Obtain $\underline{\boldsymbol{Q}}^{\mathcal{A}}$,  $\overline{\boldsymbol{Q}}^{\mathcal{A}}$ from Algorithm 1.
			\STATE 	Obtain $\underline{\boldsymbol{S}}^{\mathcal{A}}$,  $\overline{\boldsymbol{S}}^{\mathcal{A}}$ based on $\underline{\boldsymbol{Q}}^{\mathcal{A}}$ and   $\overline{\boldsymbol{Q}}^{\mathcal{A}}$ according to Algorithms 3 and 4 in \cite{stb2}.
			\STATE Solve $\pi_{u,i}$ and $\pi_{l,i}$ for $i\in \{0,1,\cdots,\alpha\}$ through Eqs. \eqref{upperpi} and \eqref{lowerpi}. Derive $\Psi_{\alpha}^u$ and $\Psi_{\alpha}^l$.
			\STATE $\varepsilon^l(j)$ for $j\in\{1,\cdots,\alpha\}$ is approximated as $
				\varepsilon^{l}(j)=1-\sum_{i=0}^{j-1}\pi_{l,i}$.
			\STATE $\varepsilon^u(j)$ for $j\in\{1,\cdots,\alpha\}$ is approximated as $\varepsilon^{u}(j)=1-\sum_{i=0}^{j-1}\pi_{u,i}$.
		\STATE Calculate $\theta_{K-i+1}$, $i=\{1,2,\cdots, \ell\}$ by solving the following equation through binary search
		\begin{equation}\label{thetaequ1}
			{\rm EC}_{K-i+1}(\theta_{K-i+1})={\rm EB}(\theta_{K-i+1}).
		\end{equation}
		\FOR{$k\in\{K-\ell+1,K-\ell+2,\cdots,K\}$}
		\FOR{$j\in\{\zeta_{k}+1,\cdots,\zeta_{k+1}\}$}
			\STATE Calculate $\varepsilon^{u}(j)$ as 
			\begin{equation}
				\varepsilon^{u}(j)\approx \varepsilon^{u}(\zeta_k) e^{-\theta_{k}(j-\zeta_k)}.
			\end{equation}
			\STATE Calculate $\varepsilon^{l}(j)$ as 
			\begin{equation}
				\varepsilon^{l}(j)\approx \varepsilon^{l}(\zeta_k)e^{-\theta_{k}(j-\zeta_k)}.
			\end{equation}
		\ENDFOR
		\ENDFOR
			\RETURN 
			$\varepsilon^{u}(j)$ and $\varepsilon^{l}(j)$, $j\in\{0,1,\cdots\}$.
	\end{algorithmic}}
\end{algorithm}

In Theorem~\ref{thm2}, we demonstrated that a censored Markov chain can approximate the decay rate of the original Markov chain following LDT with a bounded error. Next, we focus on approximating $\{\pi_i, i=\alpha+1, \cdots\}$ in the LQL regime. For this purpose, we introduce another truncated version of the original Markov chain. 

We define a new Markov chain $\{\hat{q}[n]\}$ that captures the system state for values $q[n] \in \{\alpha+1, \alpha+2, \cdots\}$. To simplify the analysis, we assume that whenever the queue experiences underflow, i.e., when the condition $\hat{q}[n] - \Xi_{\hat{q}[n]}^h + a[n+1] < \alpha + 1$ is satisfied, the state of the queue is reset to $\alpha + 1$. This assumption ensures that $\{\hat{q}[n]\}$ does not encounter underflow. Under this setup, we can utilize LDT to approximate the stationary distribution of the truncated chain $\{\hat{q}[n]\}$, denoted by $\{\hat{\pi}_i, i=\alpha+1, \alpha+2, \cdots\}$.

To overcome the difficulties in calculating the EC, the value of $\alpha$ can be chosen to equal the threshold of the cross-layer policy. For instance, setting $\alpha = \zeta_K$ degenerates the cross-layer policy into $\Lambda_K(h)$ on $[\zeta_K,+\infty)$. The EC of this policy can be computed efficiently as it does not relate to the queue state. Using this segmentation, the EC is given by \cite{ec}: 
\begin{equation}\label{ec}
	\begin{aligned}
		{\rm EC}_K(\theta_K)=&-\lim_{N\to\infty}\frac{1}{N\theta_K}\ln \mathbb{E}\left\{e^{-\theta_K\sum_{n=1}^Ns[n]}\right\}\\
		=&-\frac{1}{\theta_K}\ln \int_0^{\infty} e^{-\theta_K \Lambda_K(x)}f(x){\rm d}x.
	\end{aligned}
\end{equation}

Similarly, the EB of the arrival process is given by \cite{eb}:
\begin{equation}
	\label{eb}
	\begin{aligned}
		{\rm EB}(\theta_K)&=\lim_{n\to\infty}\frac{1}{N\theta_K}\ln\mathbb{E}\left\{e^{\theta_K \sum_{n=1}^N a[n]}\right\}\\
		&=\frac{1}{\theta_K}\ln \left(\sum_{k=0}^{\rm a_{\rm max}}p_ke^{\theta_K k}\right).
	\end{aligned}
\end{equation}

By solving the equation ${\rm EC}_K(\theta_K) = {\rm EB}(\theta_K)$, the parameter $\theta_K$ can be determined efficiently using a binary search method \cite{li2021gc}. Subsequently, the complementary cumulative distribution function (CCDF) of the chain $\{\hat{q}[n]\}$ for a given threshold $j > \alpha$ can be approximated as:
\begin{equation}
	\sum_{i=j}^{\infty}\hat{\pi}_i\approx e^{-\theta_K (j-\alpha)}.
\end{equation}

Based on this analysis, the violation probability for $q_{\rm th} \in [\zeta_K, \zeta_{K+1})$ can be approximated as:
\begin{equation}
	\varepsilon(q_{\rm th})\approx \Psi_{\alpha}^le^{-\theta_K(q_{\rm th}-\alpha)}, \quad q_{\rm th}\geq \alpha,
\end{equation}
and 
\begin{equation}
\varepsilon(q_{\rm th})\approx\Psi_{\alpha}^ue^{-\theta_K(q_{\rm th}-\alpha)}, \quad q_{\rm th}\geq \alpha,
\end{equation}

The above analysis focuses on the interval $[\zeta_K, \zeta_{K+1})$. By adjusting the interval, this approach can be applied on $\ell$ segments, leading to a piecewise result. Specifically, we set $\alpha = \zeta_{K-\ell+1}$ so that the states $\{0, 1, \cdots, \zeta_{K-\ell+1}\}$ belong to the SQL regime. For the LQL regime, the analysis is conducted over $\ell$ segments, i.e., $[\zeta_{K-\ell+1}, \zeta_{K-\ell+2}), \cdots, [\zeta_K, \zeta_{K+1})$. Within each segment, the QoS exponent $\theta_{K-i+1}$ can be obtained by solving ${\rm EC}_{K-i+1}(\theta_{K-i+1}) = {\rm EB}(\theta_{K-i+1})$, where $i \in \{1, 2, \cdots, \ell\}$.

Based on these results and the analysis presented in Section~\ref{smallq}, we propose an algorithm to approximate the tail distribution of $\{q[n]\}$ by utilizing stochastic bounds of the censored Markov chain and a piecewise LDT method. Algorithm~\ref{alg2} details this procedure. In Algorithm~\ref{alg2}, $\varepsilon^l(q_{\rm th})$ and $\varepsilon^u(q_{\rm th})$ represent the approximations for $\varepsilon(q_{\rm th})$, calculated based on $\underline{\boldsymbol{S}}^{\mathcal{A}}$ and $\overline{\boldsymbol{S}}^{\mathcal{A}}$, respectively.

EVT can also be used for characterizing the more general case of the tail distribution of queue length, e.g., without maximum power constraint \cite{li2023}, and the largest queue length of massive independent queues \cite{liu2019}. We will also analyze the accumulated error between the decay rate of the censored Markov chain and the original chain following EVT in the following two theorems.

\begin{theorem}
	\label{thm3}
	Assume $\{q[n]\}$ satisfies Pickands-Balkema-De Haan Theorem, i.e.,
	\begin{equation}\label{thm3e1}
		\begin{aligned}
		\lim_{d\to\infty}& \Pr\left\{q[n]>x \,| \, q[n]>d\right\}\approx\\
		& \quad \quad \quad \quad \begin{cases}
			\left(1 + \frac{\tilde{\xi} y}{\tilde{\sigma}}\right)^{-\frac{1}{\tilde{\xi}}}, & \tilde{\xi} > 0, \\
			e^{-\frac{y}{\tilde{\sigma}}}, & \tilde{\xi} = 0,
		\end{cases}.
		\end{aligned}
	\end{equation}
\begin{itemize}
	\item 	The corresponding $\vartheta(\alpha)$ is unbounded with $\tilde{\xi}>0$.
	\item  For $\tilde{\xi}=0$, we have
	\begin{equation}
		\frac{1}{e^{\frac{1}{\tilde{\sigma}}}-1}\leq \lim_{\alpha\to+\infty}\vartheta(\alpha)\leq \frac{1}{(e^{\frac{1}{\tilde{\sigma}}}-1)(1-e^{-\frac{1}{\tilde{\sigma}} })}.
	\end{equation}
\end{itemize}
\end{theorem}

\begin{IEEEproof}
	See Appendix D.
\end{IEEEproof}

\begin{theorem}
	\label{thm4}
	Assume $\{q[n]\}$ satisfies Fisher-Tippett-Gnedenko Theorem, i.e., 
	\begin{equation}\label{ttt000}
		\Pr\{q[n]>y\}\approx 	\begin{cases}
			1-e^{-\left(1+\frac{\xi(y-\mu)}{\sigma}\right)^{-\frac{1}{\xi}}}, & \xi \neq 0, \\
			1-e^{-e^{-\frac{y-\mu}{\sigma}}}, & \xi = 0,
		\end{cases}.
	\end{equation}
	\begin{itemize}
		\item  For $\xi>1$, $\vartheta(\alpha)$ is unbounded.
		\item The corresponding $\vartheta(\alpha)$ with $\xi=1$ satisfies
		\begin{equation}
			0\leq \lim_{\alpha\to+\infty}\vartheta(\alpha)\leq \sigma+\sigma^2.
		\end{equation}	
		\item The corresponding $\vartheta(\alpha)$ with $\xi\in(0,1)$ satisfies
		\begin{equation}
			\label{thm3e2}
			0\leq \lim_{\alpha\to+\infty}\vartheta(\alpha)\leq \frac{\sigma^2}{\xi}.
		\end{equation}	
	\item The corresponding $\vartheta(\alpha)$ with $\xi=0$ satisfies
	\begin{equation}
		0\leq \lim_{\alpha\to+\infty}\vartheta(\alpha)\leq \frac{e^{-\frac{1-\mu}{\sigma}}}{1-e^{-\frac{1}{\sigma}}}.
	\end{equation}

	\end{itemize}
\end{theorem}
\begin{IEEEproof}
	See Appendix E.
\end{IEEEproof}

Unlike the combination of the censored Markov chain and LDT, the combination of the censored Markov chain and EVT still requires data samples to estimate the parameters of GEV and GPD. To address this limitation, the importance sampling method can be employed to reduce the cost of obtaining these samples \cite{imsam}, which represents a significant direction for future research.

Combining with Theorem~\ref{thm2}, we can analyze the distribution of the maximum queue length without data samples. First, we obtain the distribution of queue length based on Algorithm~\ref{alg2}, which does not need the aid of data samples. Second, the CCDF of the maximum queue length of $N$ independent queues with the same scheduling policy can be obtained by $1-\varepsilon^N(q)$. Based on these results, we have obtained an approximate curve of CCDF, which can be used to conduct curve-fitting for the GEV or GPD distribution. As a result, Theorems~\ref{thm3} and~\ref{thm4} are then used to ensure that combining EVT and censored Markov chain leads to a bounded error for decay rate approximation.

\subsection{\R{Discussion on Scheduling Design Using the Proposed Method}}

\R{In the preceding subsections, we demonstrate that the proposed approach can effectively evaluate the QVP performance for a given communication system. In this subsection, we outline how to leverage this approach to design a scheduling policy that achieves a prescribed QVP, thereby enabling optimization of practical systems. Specifically, we formulate an optimization problem whose objective is to minimize the average transmit power subject to a QVP constraint.}

\R{For simplicity, we assume that $q_{\rm th}=\zeta_{\varsigma}$, where $\varsigma \in  \{1,\cdots, K\}$. We partition the queue-length state space into outer states indexed by $k \in \{1,\dots,K+1\}$, where each outer state corresponds to the interval $[\zeta_{k-1}, \zeta_k)$. We denote the stationary probability of the $k$-th outer state by $\pi_k^{\rm I}$. Given the threshold $\{\zeta_1,\cdots, \zeta_K\}$~\footnote{\R{Note that the choice of $K$ impacts the signaling overhead. A larger $K$ enables finer‐grained transmission control but incurs higher overhead, since the base station must receive more precise queue‐state information. Determining the optimal value of $K$ is therefore an important topic for future investigation.}}, the optimization problem is formulated as}
\begin{subequations}\label{opj}
	\R{\begin{align}
			\min & \sum_{i=1}^{K+1} \pi_{i}^{\rm I} \bar{P}_i \nonumber \\
			\mbox{s.t.}\quad
			&  \!\!\!\! \sum_{i=1}^{K+1} \pi_i^{\rm I}\left(\sum_{j=\varsigma+1}^{K+1}\beta_{i,j}\right)\leq \varepsilon_{\rm th}, \label{cons1}\\
			& \!\!\!\!\pi_j^{\rm I}\!=\! \sum_{i\leq j}\pi_i^{\rm I}\beta_{i,j}+\sum_{i> j}\pi_i^{\rm I}\varrho_{i,j}, \, j\in\{1,\cdots,K+1\}, \label{cons2}\\
			&\!\!\!\!\sum_{i=1}^{K+1} \pi_i^{\rm I}=1, \label{cons3}\\
			&\!\!\!\!\beta_{i,j}\!=\!\Omega^{\rm B}_{i,j}\left(\!\bar{P}_i\!\right), \, i\!\leq\! j, \text{ and } i,j \in \left\{1,\cdots,K+1\right\}, \label{cons4}\\
			& \!\!\!\!\varrho_{i,j}\!=\!\Omega^{\rm F}_{i,j}\left(\!\bar{P}_i\!\right),\,  i\!>\! j, \text{ and } i,j \in \left\{1,\cdots,K+1\right\}, \label{cons5}
	\end{align}}
\end{subequations}
\R{where Eq. \eqref{cons1} is the constraint on QVP. Eqs. \eqref{cons2} and \eqref{cons3} are the steady-state balance equations. $\Omega_{i,j}^{\rm B}(\cdot)$ and $\Omega_{i,j}^{\rm F}(\cdot)$ denote the backward transition probability and forward transition probability, respectively, induced by the scheduling policy $\Lambda_i(h)$ over the interval $[\zeta_i,\zeta_{i+1})$.}

\begin{figure}[t]
	\centerline{\includegraphics[width=9cm]{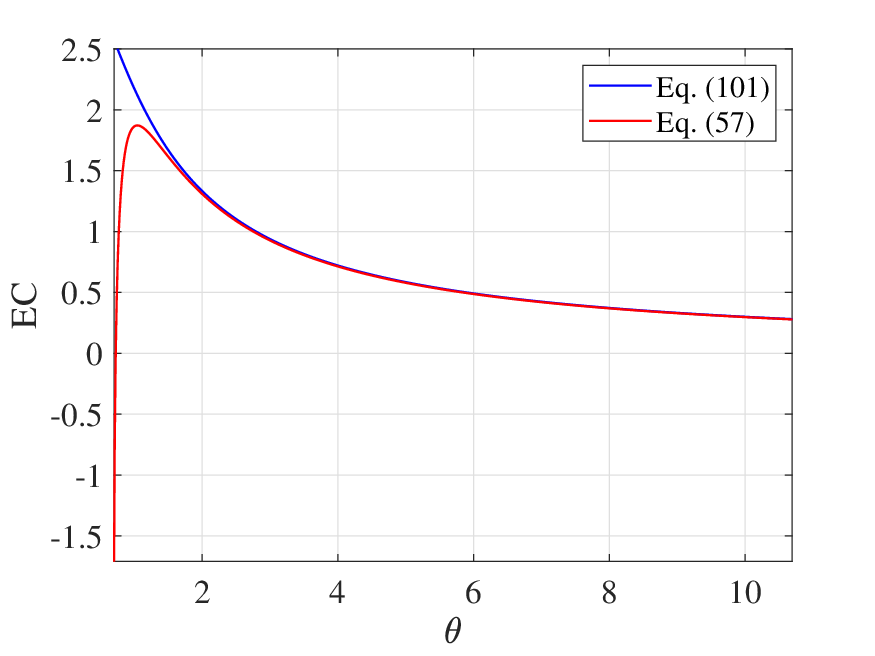}}
	\caption{Comparison between Eq. \eqref{ecexpression} and Eq. \eqref{ecapprox}.}
	\label{eccompare}
\end{figure}

\R{As described above, we find that $\Omega_{i,j}^{\rm B}(\cdot)$ and $\Omega^{\rm F}_{i,j}(\cdot)$ depend on the scheduling policy $\Lambda_i(h)$. Once a scheduling policy $\Lambda_i(h)$ is given, we have to solve a steady-state equation for the inner states in $[\zeta_i,\zeta_{i+1})$ to determine $\Omega_{i,j}^{\rm B}(\cdot)$ and $\Omega^{\rm F}_{i,j}(\cdot)$, which renders the optimization in Eq. \eqref{opj} intractable in its raw form. To address this difficulty, the proposed method in this paper can be applied as follows:}
\R{	\begin{enumerate}
		\item Select $\alpha$ according to the computational capabilities of the system.
		\item Approximate the QVP in the SQL regime $[0,\alpha]$ using the method from Section III-A.
		\item For each outer interval $[\zeta_i,\zeta_{i+1})$ in the LQL regime, fix $\Lambda_i(h)$ and derive the EC via Eq. \eqref{ec}. Since $\Lambda_i(h)$ depends only on the CSI in $[\zeta_i,\zeta_{i+1})$, EC can be obtained through the analytical expression.
		\item Solve for $\theta_i$ using the equation ${\rm EC}_i(\theta_i) = {\rm EB}(\theta_i)$ by binary search.  
		\item Repeat steps 2–4 (corresponding to steps 6–12 of Algorithm 2) to obtain an efficient approximation of the QVP, which in turn informs the design of $\Lambda_i(h)$.
\end{enumerate}}

\R{Due to space limitations, the detailed derivations of the optimization procedure are left for future work.}

\begin{figure*}[b]
	\hrulefill
	\setcounter{equation}{50}
	\begin{equation}\label{tp1}
		\begin{aligned}
		\!\!\!\!\!	\mathcal{F}_{i,j}^{(1)}=\left\{x|x\geq\frac{ N_0VA}{2T}\left(\left\lfloor \frac{i\kappa_q}{\delta}\right\rfloor \delta+\lambda\right)^{-1}\right\}&\cap\left\{x|x<\frac{N_0VA}{2T}\left(\left\lfloor \frac{i\kappa_q}{\delta}\right\rfloor\delta+\lambda\right)^{-1}\min\left\{2^{\frac{\kappa_q A\left(i-j+1+\frac{\lambda}{\kappa_q}\right)}{BT}},2^{\frac{\kappa_q A(i+1)}{BT}}\right\}\right\} \\
			&\cap\left\{x|x\geq \frac{N_0VA}{2T}\left(\left\lfloor \frac{i\kappa_q}{\delta}\right\rfloor\delta+\lambda\right)^{-1}2^{\frac{\kappa_q A\left(i-j+\frac{\lambda}{\kappa_q}\right)}{BT}}\right\}.
		\end{aligned}
	\end{equation}
	
	\begin{equation}\label{tp2}
		\mathcal{F}_{i,j}^{(2)}=\left\{x|x< \frac{N_0VA}{2T}\left(\left\lfloor \frac{i\kappa_q }{\delta}\right\rfloor\delta+\lambda\right)^{-1}\mathbb{I}\left\{-1<i-j+\frac{\lambda}{\kappa_q}\leq 0\right\}\right\}\cap \left\{x| x\geq 0\right\}.
	\end{equation}
		\setcounter{equation}{54}
	\begin{equation}\label{thetaequ}
		\begin{aligned}
				\lambda=-\frac{1}{\theta_K} \ln \left[ 1-\exp\left( -\frac{N_0VA}{2T\left(\omega \delta +\lambda\right)}  \right) +\left(  \frac{N_0VA}{2T(\omega \delta+\lambda)}  \right)^{\frac{BT}{A}\theta_K\log_2 e }\Gamma\left(1-\frac{BT}{A}\theta_K\log_2 e, \frac{N_0VA}{2T(\omega\delta+\lambda)}\right)\right] .
			\end{aligned}
	\end{equation}
	
\end{figure*}

\section{Applying the Tractable Method for Analyzing A Lyapunov-Drift-Based Cross-Layer Scheduling}

In this section, we demonstrate how the proposed two-stage method can be applied to analyze the tail distributions of wireless systems adopting buffer-aware scheduling policies. As an example, we consider a Lyapunov-drift-based cross-layer scheduling policy. For simplicity, we assume that $\kappa_q$ remains two constants in the SQL and LQL regimes and $\frac{\alpha}{\kappa_q}\in \mathbb{Z}^+$.

\subsection{Introduction to the Lyapunov-drift-Based Cross-Layer Scheduling}

The considered Lyapunov-drift-based scheduling policy dynamically adjusts the transmission power based on QSI and CSI. Thus, it belongs to buffer-aware scheduling policy. For detailed implementation, readers are referred to \cite{Xie2022}. Here, we assume a system with a constant arrival rate $a[n] = \lambda$.\footnote{This policy is also applicable to systems with stochastic arrivals, and the proposed analysis extends naturally to the stochastic arrival case.} Given QSI and CSI, the transmission power is determined by \cite{Xie2022}
\setcounter{equation}{47}
\begin{equation}\label{lya}
	P[n]=\left(\frac{2BT}{VA}\left\lfloor \frac{q[n]}{\delta}\right\rfloor \delta +\frac{2BT}{VA}\lambda-\frac{N_0B}{|h[n]|^2}\right)^{+},
\end{equation}
where $V$ is a hyper-parameter controlling the Lyapunov drift, and $\delta \in \mathbb{N}_+$ denotes the scheduling granularity of the queue length. With the obtained $P[n]$, the transmission rate is thus determined by Eq. \eqref{cap}. According to Eqs. \eqref{x3} and \eqref{x4}, the queue length evolves as follows:
\begin{equation}
	q[n+1]=\left( q[n]-\left\lfloor\frac{C[n]}{\kappa_qA}\right \rfloor\kappa_q\right)^+ + \lambda.
\end{equation}

Our goal is to obtain $\Pr\{q[n]\geq m\}$, where $m=0,1,\cdots$.  In Section IV-B, we will show how to obtain $Q_{\mathcal{A},\mathcal{A}}$ so that we can derive the stochastic upper and lower bounds of $\pi_i^{\alpha}$, where $i=0,1,\cdots,\alpha$. In Section IV-C, we will present the analysis of utilizing the piecewise LDT to approximate the tail distribution in the LQL regime.

\subsection{Analysis in the SQL Regime}

In this subsection, we derive the stochastic bounds for the censored Markov chain in the SQL regime. To achieve this, we first compute $Q_{\mathcal{A},\mathcal{A}}$.

\begin{theorem}
	\label{thm5}
For a communication system employing the scheduling policy in Eq. \eqref{lya}, the $(i,j)$-th element of $Q_{\mathcal{A},\mathcal{A}}$ is given by:
	\begin{equation}\label{ijelements}
		Q_{\mathcal{A},\mathcal{A}}^{i,j}=\int_{\mathcal{F}_{i,j}} \mathbb{I}\left\{j\leq i+\lambda \right\}f(x) {\rm d}x + \int_{\mathcal{G}_{i,j}} \mathbb{I}\{j=\lambda\}f(x){\rm d}x,
	\end{equation}
	where $i,j \in \{0,1, \cdots, \frac{\alpha}{\kappa_q}\}$, $\mathcal{F}_{i,j} = \mathcal{F}_{i,j}^{(1)} \cup \mathcal{F}_{i,j}^{(2)}$, with $\mathcal{F}_{i,j}^{(1)}$ and $\mathcal{F}_{i,j}^{(2)}$ defined in Eqs. \eqref{tp1} and \eqref{tp2} at the bottom of this page, respectively. $\mathcal{G}_{i,j}$ is defined as:
	\setcounter{equation}{52}
	\begin{equation}\label{tp3}
		\mathcal{G}_{i,j}=\left\{x|x\geq \frac{N_0VA}{2T}\left(\left\lfloor \frac{i\kappa_q }{\delta}\right\rfloor\delta+\lambda\right)^{-1}2^{\frac{\kappa_q A(i+1)}{BT}}\right\}.
	\end{equation}
\end{theorem}

\begin{IEEEproof}
	See Appendix F.
\end{IEEEproof}

Using Eq. \eqref{ijelements}, $Q_{\mathcal{A},\mathcal{A}}$ can be explicitly computed. Consequently, the proposed analysis in Section III can be applied to approximate and derive the stochastic bounds for the tail distribution in the SQL regime.

\begin{table}[t]
	\small 
	\centering 
	\caption{System parameters}  
	\begin{center}  
		\begin{tabular}{|c|c||c|c|} 
			\hline  
			\multicolumn{1}{|c|}{\textbf{Parameter}} & \multicolumn{1}{c||}{\textbf{Value}} & \multicolumn{1}{c|}{\textbf{Parameter}} & \multicolumn{1}{c|}{\textbf{Value}} \\ \hline  
			$N_0$ & -174 dBm/Hz & $\mathbb{E}\{|h[n]|^2\}$ & $5\times 10^{-15.4}$ \\ \hline
			$B$ & 500 KHz     &    $T$   & 2 ms \\ \hline
			$V$ & 2 & $A$  &  $10^3$ bit \\ \hline  
			$\lambda$ & $1$ &  $\varkappa$ & $10^9$ \\ \hline
			$\kappa_s$ & $1$ &  $\kappa_l$ & $10^{-3}$ \\ \hline
		\end{tabular}  
	\end{center}  
\end{table}

\subsection{Analysis in the LQL Regime}

In this subsection, we demonstrate the application of Algorithm 2 to a system adopting the Lyapunov-drift-based scheduling policy. In Section IV-A, we obtained the corresponding $\underline{Q}^{\mathcal{A}}$ and $\overline{Q}^{\mathcal{A}}$. Thus, Steps 1-5 of Algorithm 2 can be executed. Next, we need to solve the equation shown in Eq. \eqref{thetaequ1}. To clearly illustrate the use of Algorithm 2, we take the Rayleigh fading channel, i.e., $f(x) = e^{-x}$ for $x > 0$, as an example.

\R{Given $a[n] = \lambda$, we have \cite{eb}:}
\setcounter{equation}{53}
\begin{equation}\label{ebe}
	\rm{EB}\left(\theta\right)=\lambda.
\end{equation}

Since the Lyapunov-drift-based scheduling policy is quantized with granularity $\delta$, we assume the scheduling threshold $K$ equals $\omega \delta$, where $\omega \in \mathbb{N}_+$. Substituting Eqs. \eqref{ebe} and \eqref{ec} into Eq. \eqref{thetaequ1}, we can numerically solve $\theta_K$. To reduce computation complexity, $\theta_K$ can be approximated with a closed-form expression. In Theorem 6, we present this closed-form approximation under the assumption of large $\omega$.

\begin{theorem}
\label{thm6}
For large $\omega$, the QoS exponent $\theta_K$ can be approximated by solving Eq. \eqref{thetaequ}, provided at the bottom of this page.
\end{theorem}

\begin{IEEEproof}
	See Appendix G.
\end{IEEEproof}

In Theorem \ref{thm6}, we derive the equation for solving $\theta_K$. Since ${\rm EC}_K(\theta_K)$ is an increasing function with respect to $\theta_K$ \cite{ec}, this equation can be efficiently solved via binary search. Moreover, Eq. \eqref{thetaequ} can be further simplified, as summarized in Corollary \ref{coro1}. For simplicity, let $\psi = \frac{N_0VA}{2T(\omega \delta + \lambda)}$.

\begin{coro}\label{coro1}
 For large $\omega$, the solution of Eq. \eqref{thetaequ} can be approximated by solving the following equation:
 \setcounter{equation}{55}
 \begin{equation}\label{appxequ}
 	\theta_K-\frac{1}{\lambda}\ln\left(1-\frac{A}{BT\theta_K\log_2e}\right)= -\frac{1}{\lambda}\ln \psi.
 \end{equation}	
\end{coro}

\begin{IEEEproof} 
	See Appendix H.
\end{IEEEproof}

\begin{rem}
	\emph{In the proof of Corollary \ref{coro1}, we have mentioned that  $\theta_K\to +\infty$ when $\omega\to+\infty$. Therefore, under the assumption that $\omega$ is large, the dominant term of the left-hand-side (LHS) of Eq. \eqref{appxequ} is $\theta_K$. Therefore, a loose approximation of $\theta_K$ is}
	\begin{equation}
		\theta_K\approx -\frac{1}{\lambda}\ln\psi.
	\end{equation}
	\emph{Note that this approximation only has satisfying performance when $\omega \to +\infty$, which leads to $\theta_K\to+\infty$. The accuracy of this approximation can be shown in Fig.~\ref{eccompare}. For a stricter approximation, we can utilize the property that $x-a\ln(1-\frac{b}{x})$ is increasing over $[c,+\infty)$, where $a$, $b$, and $c$ are positive constants. Since $\theta_K$ is large, we can treat the LHS of Eq. \eqref{appxequ} as an increasing function of $\theta_K$. Then we can adopt the binary search to obtain the solution of Eq. \eqref{appxequ}. }
\end{rem}

\begin{figure}[t]
	\centerline{\includegraphics[width=9 cm]{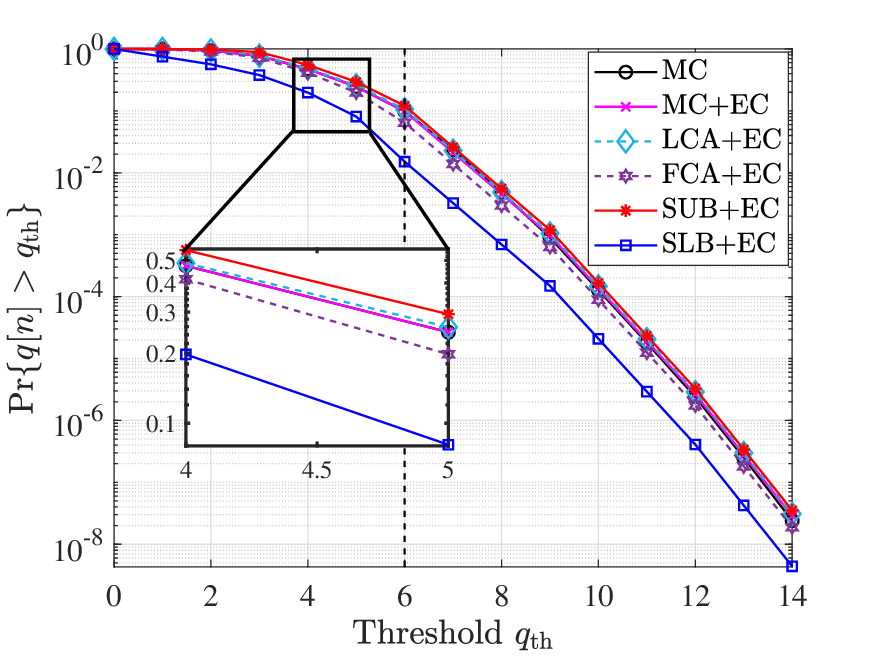}}
	\caption{QVP versus threshold with $V=2$, $\delta=3$, and $\alpha=6$.}
	\label{fig2}
\end{figure}

\begin{rem}
\emph{Before conducting binary search, the search interval must be determined. The upper bound can be simply set as $-\frac{1}{\lambda} \ln \psi$. The lower bound depends on parameters such as $\lambda$, $A$, $B$, and $T$. When $\theta_K$ is large, the logarithmic term becomes negligible. Thus, the lower bound can be approximated as $-\frac{\upsilon}{\lambda} \ln \psi$, where $\upsilon \in (0, 1)$ and typically approaches 1. The theoretical analysis of the lower bound is left as future work.}
\end{rem}

\section{Simulation and Numerical Results}

This section presents the simulation and numerical results corresponding to the conclusions drawn in Sections III and IV. The simulation results are obtained using the MC method and averaged over $\varkappa$ time slots. For simplicity, we set $\kappa_q = \kappa_s$ for $q \in [0, \alpha]$ and $\kappa_q = \kappa_l$ for $q > \alpha$. Unless specified otherwise, the parameters used in the simulations and numerical calculations are summarized in Table III.  \R{We consider a non-line-of-sight urban micro-cell scenario as proposed by 3GPP in \cite{3GPP2}. The parameters $B$ and $T$ are selected based on the typical values of coherence bandwidth and coherence time as suggested in \cite{3GPP2}. The carrier frequency is set to $f_c=3.5$ GHz, which is commonly adopted in practical scenarios \cite{3GPP0,3GPP1}. Moreover, we assume that the distance between the transmitter and receiver is around $D=1.5$ km. According to \cite{3GPP2}, the value of $|h[n]|^2$ is given by}
\begin{equation}\label{x7}
	\!\!\!\! \R{	10\log_{10} \!  \mathbb{E}\left\{|h[n]|^2\right\} \!=\! -36.7\log_{10} \! D \!- \!22.7 \! - \! 26\log_{10}f_c.}  
\end{equation}
\R{By substituting the specified value of $f_c$ and $D$ into Eq. \eqref{x7}, we obtain the corresponding result as listed in Table III. Moreover, we set $\kappa_{s}=1$ to enable packet-level control in the SQL regime, while we set $\kappa_{l}=10^{-3}$ to facilitate bit-level control in the LQL regime. The control parameter $V$ is varied in the subsequent analysis to demonstrate its impact. It is worth noting that all parameters can be adjusted to reflect specific practical deployment scenarios.}

\begin{figure}[t]
	\centerline{\includegraphics[width=9 cm]{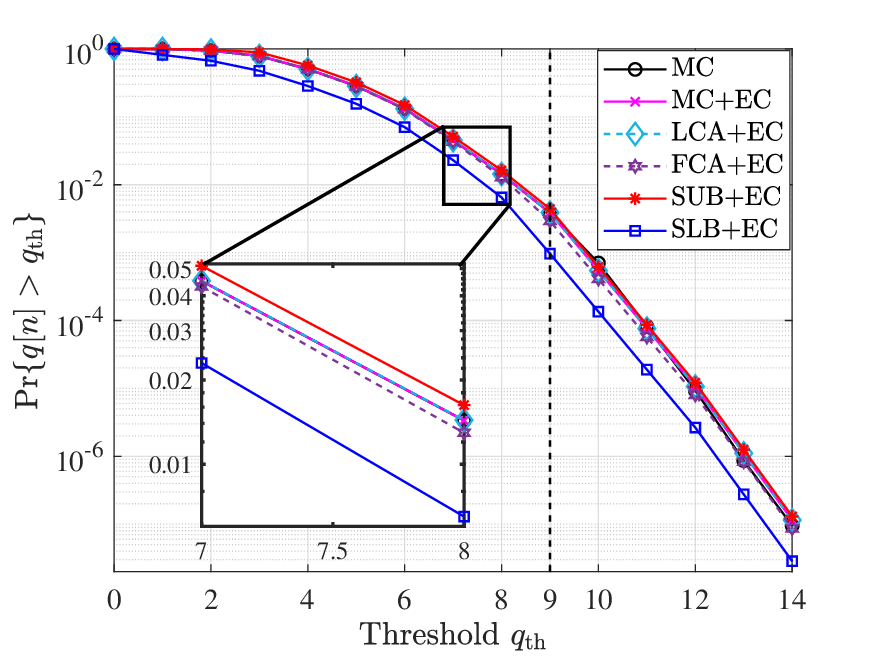}}
	\caption{QVP versus threshold with $V=2$, $\delta=3$, and $\alpha=9$.}
	\label{fig3}
\end{figure}

\begin{figure}[t]
	\centerline{\includegraphics[width=9 cm]{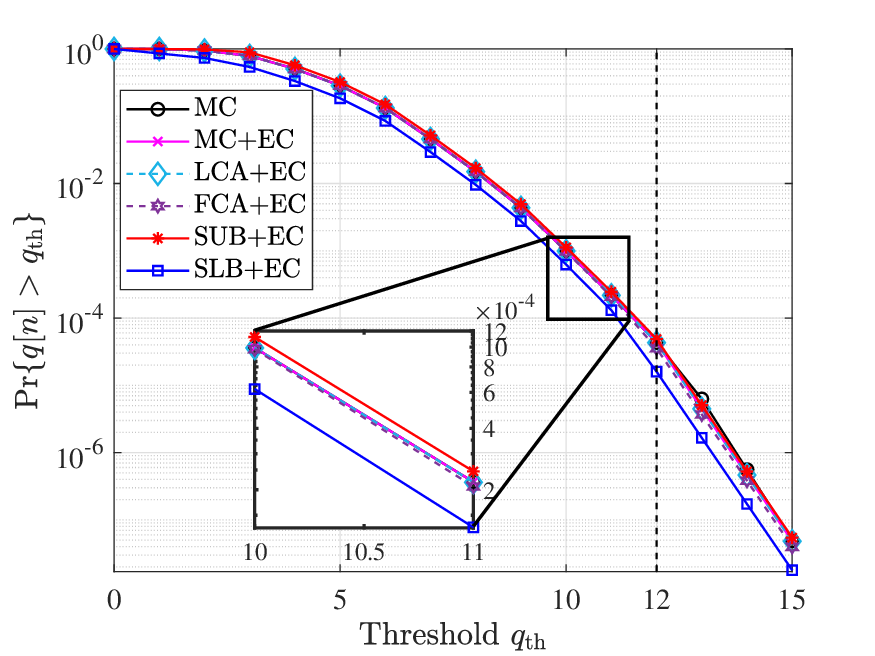}}
	\caption{QVP versus threshold with $V=2$, $\delta=3$, and $\alpha=12$.}
	\label{fig4}
\end{figure}

\begin{figure}[t]
	\centerline{\includegraphics[width=9 cm]{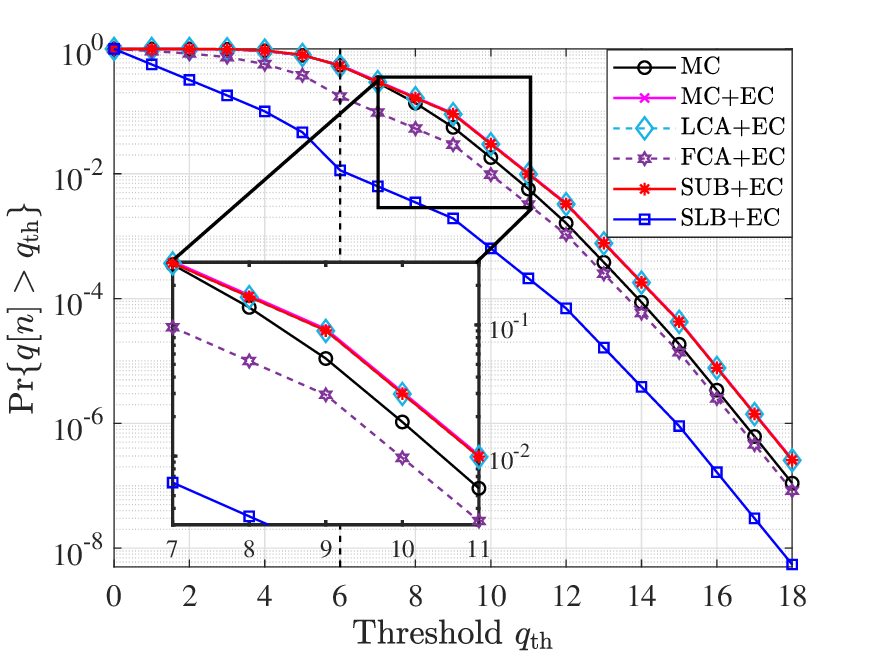}}
	\caption{QVP versus threshold with $V=4$, $\delta=3$, and $\alpha=6$.}
	\label{fig5}
\end{figure}

\begin{figure}[t]
	\centerline{\includegraphics[width=9 cm]{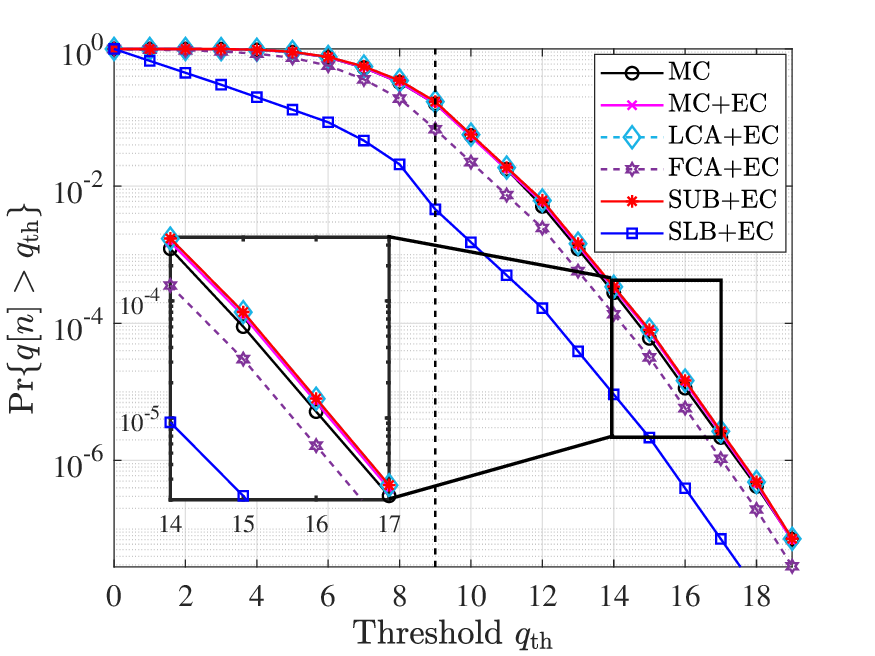}}
	\caption{QVP versus threshold with $V=4$, $\delta=3$, and $\alpha=9$.}
	\label{fig6}
\end{figure}

\begin{figure}[t]
	\centerline{\includegraphics[width=9 cm]{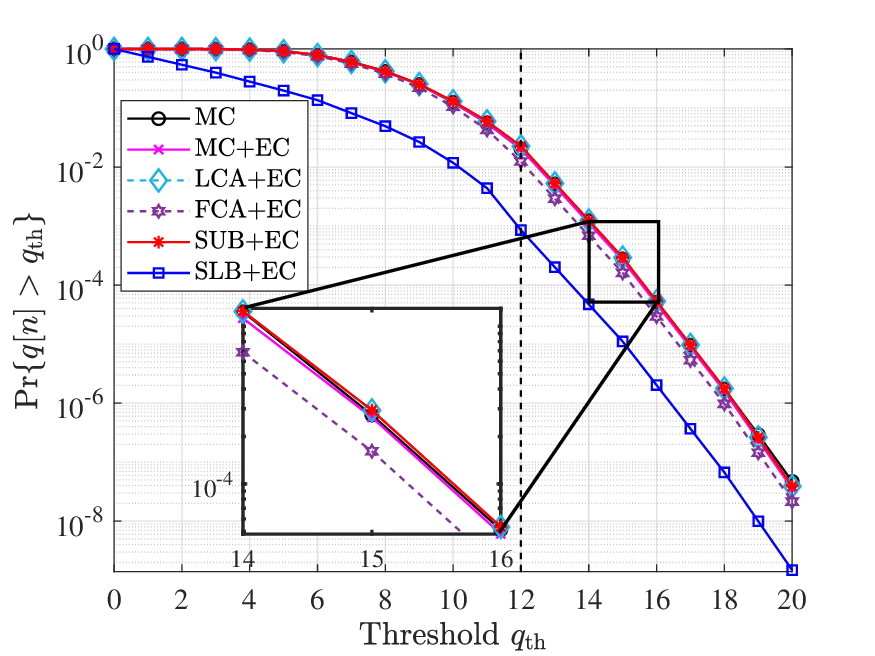}}
	\caption{QVP versus threshold with $V=4$, $\delta=3$, and $\alpha=12$.}
	\label{fig7}
\end{figure}

Before delving into the results, we explain the legends and auxiliary lines in Figs. ~\ref{fig2}--\ref{fig7}. These figures divide the queue length into two regions, with the black dotted line representing $\alpha$, which is the boundary between the SQL and LQL regimes. The legend details are given as follows:
	\begin{itemize}
		\item \texttt{MC}: Results obtained from MC simulations.
		\item \texttt{A+B}: Results using Method A in the SQL regime and Method B in the LQL regime.
		\item \texttt{EC}: Results using the piecewise EC (i.e., piecewise LDT) method presented in Algorithm~2.
		\item \texttt{LCA} and \texttt{FCA}: Last-column and first-column augmentations in Algorithm~1, respectively.
		\item \texttt{SUB} and \texttt{SLB}: Stochastic upper and lower bounds derived from Algorithms 3 and 4 in \cite{stb2}.
	\end{itemize}
	Note that, to show the accuracy of the piecewise-EC based approximation in the LQL regime, the curves with legend `MC+EC' set the QVP equal to that of curves with legend `MC' in the SQL regime. Therefore, the MC curves coincide with the `MC+EC' curves in the SQL regime.

Figs.~\ref{fig2}--\ref{fig4} present QVP versus the threshold with $V=2$, $\delta=3$, and $\alpha \in \{6,9,12\}$. Similarly, Figs.~\ref{fig5}--\ref{fig7} show results for $V=4$, $\delta=3$, and $\alpha \in \{6,9,12\}$. The selection of $\alpha$ should reflect the computational capabilities of the system under consideration. In this work, we set $\alpha$ to be a positive integer multiple of $\delta$. In the SQL regime, it is observed that the curve labeled `SUB+EC' serves as an upper bound for the actual QVP when $q_{\rm th} < \alpha$, whereas the curve labeled `SLB+EC' acts as a lower bound under the same condition. This behavior arises from the censored chain, which serves as the best approximation of the original chain in the SQL regime, resulting in a small discrepancy between the two. Consequently, the stochastic upper and lower bounds derived from the censored Markov chain are also effective bounds for the original chain in this regime. This makes the SLB a suitable conservative approximation for QVP in practical applications. Furthermore, as $\alpha$ increases, the gaps between the SUB and the actual QVP, as well as between the LCA and the actual QVP, progressively diminish. This trend is reasonable because a larger $\alpha$ reduces the probability of $q[n] > \alpha$, thereby narrowing the difference between the truncated chain and the original chain, which enhances the approximation accuracy of the LCA. Since the SUB is derived from the LCA, it also exhibits improved accuracy with increasing $\alpha$. Comparing results in Figs.~\ref{fig2}--\ref{fig4} and Figs.~\ref{fig5}--\ref{fig7}, it is evident that the gap between the SLB and the original chain is more pronounced for $V=4$ than for $V=2$. A similar trend is observed for the FCA. This can be attributed to the fact that a larger $V$ corresponds to reduced power availability for scheduling, which increases $\Pr\{q[n] > \alpha\}$ for a given $\alpha$. The FCA approximation assumes that the buffer is emptied when $q[n] > \alpha$, whereas the original chain is less likely to empty the buffer under the same conditions when $V$ is large. As a result, the gap between the FCA and the actual chain becomes more significant with larger $V$. Since the SLB is based on the FCA, it exhibits a similar behavior. Additionally, we also find that the gap between the SUB and SLB becomes smaller with the increase of $\alpha$, which is an important future work to prove it theoretically.

Next, we discuss the results in the LQL regime. It is observed that the curves labeled `MC+EC' closely align with the MC results in this regime, demonstrating that the piecewise-LDT-based approximation, as presented in Alg.~\ref{alg2}, provides high accuracy for the QVP in the LQL regime. The piecewise property is clearly evident in Fig.~\ref{fig5}. Across all the figures, the `LCA+EC' method emerges as a reliable approximation for the considered Lyapunov-drift-based scheduling policy. However, the performance of the LCA-based approximation may not always remain robust, as its accuracy depends on the transition probability of the original chain. The theoretical justification of the considered approximation methods in the SQL regime, as well as the development of a more general and accurate approximation for the QVP in the SQL regime, remain important directions for future research. Regardless of the specific approximation adopted in the SQL regime, the piecewise LDT effectively captures the decay rate of the QVP in the LQL regime, demonstrating its practical applicability.

Finally, we compare the computation time for MC simulations and the proposed approximations. All the simulations and numerical results are conducted on an Intel i9-13900KF CPU. The MC simulations required more than 1100 seconds to complete. For the approximations in the SQL regime, the most time-consuming step involves formulating $Q_{\mathcal{A},\mathcal{A}}$, which takes less than 1.5 seconds. Once $Q_{\mathcal{A},\mathcal{A}}$ is obtained, calculating the LCA, FCA, SUB, and SLB approximations requires approximately 0.003 seconds. The subsequent computation of the piecewise LDT takes less than 0.6 seconds. In total, the computation time for all approximation methods is under 2.2 seconds, which is significantly faster than MC simulations. In the context of 6G scenarios with more stringent requirements (e.g., $\Pr\{q[n]>q_{\rm th}\}<10^{-9}$), MC simulations would need to increase the order of $\varkappa$ to achieve reliable performance, resulting in impractical computation times. This underscores the advantage of the proposed approximations in achieving accurate QVP estimates with reduced computational overhead. Additionally, we also find that we can propose a hybrid approach that combines MC simulations with the piecewise-LDT-based approximation. For instance, setting $\varkappa=10^5$ allows the approximation of QVP values exceeding $10^{-4}$ within approximately 0.11 seconds. The piecewise-LDT-based method can then address the remaining parts. This hybrid strategy further reduces computation time while potentially enhancing accuracy, as the MC simulations directly approximate the actual QVP. These findings highlight the benefits of integrating numerical and theoretical methods for performance analysis.

\section{Conclusion}

In this paper, we proposed a fully analytical approach to conduct queueing analysis in wireless systems adopting buffer-aware scheduling policies. This approach is more tractable than traditional methods used for queueing analysis in systems with buffer-aware scheduling, where the service rate is inherently coupled with the queue length. Specifically, the queue length was divided into two regimes: the SQL and LQL regimes. For the SQL regime, we employed a truncated STM to approximate the QVP, leveraging first-column augmentation, last-column augmentation, and censored Markov chain-based techniques with its stochastic upper and lower bounds. We proved that for buffer-aware scheduling policies with the threshold property, last-column augmentation yields the stochastic upper bounds of the censored Markov chain. In the LQL regime, we proposed a piecewise analytical method that segments the LQL to apply LDT or EVT-based approximations. We theoretically established that the accumulated error between the actual QVP and the proposed approximation is bounded. Moreover, we derived closed-form expressions for the accumulated errors under various SQL and LQL approximation combinations. Numerical and simulation results validate the effectiveness of the proposed approximations and reveals the potential of integrating MC simulations with the proposed piecewise method as a new computational probability method. Important future research directions include extending the analysis to LVP, designing tighter bounds or more precise approximations for the QVP in the SQL regime, and theoretically analyzing the performance of integration of the proposed method with MC simulations.

\appendices

\section{Proof of Theorem~\ref{thm2}}
	By substituting Eq. \eqref{cenchainstprob} into \eqref{l1er}, we obtain
\begin{equation}\label{orierror}
	\begin{aligned}
			\vartheta(\alpha)&=\ln\left(  \frac{ \prod_{k=0}^{\alpha-1}\left(1-\sum_{i=0}^k\pi_i\right)   }{ \prod_{k=0}^\alpha \left(1-\frac{1}{1-\varepsilon(\alpha)}\sum_{i=0}^k\pi_i\right)   } \right) \\
		&=\alpha\ln\left(1-\varepsilon(\alpha)\right)+\ln\left(  \frac{ \prod_{k=0}^{\alpha-1}\left(1-\sum_{i=0}^k\pi_i\right)   }{ \prod_{k=0}^{\alpha-1} \left(1-\varepsilon(\alpha)-\sum_{i=0}^k\pi_i\right)   } \right).
	\end{aligned}	
\end{equation}

For simplicity of notation, we define $\iota_1(\alpha)=\alpha\ln\left(1-\varepsilon(\alpha)\right)$ and $\iota_2(\alpha)=\ln\left(  \frac{ \prod_{k=0}^{\alpha-1}\left(1-\sum_{i=0}^k\pi_i\right)   }{ \prod_{k=0}^{\alpha-1} \left(1-\varepsilon(\alpha)-\sum_{i=0}^k\pi_i\right)   } \right)$. We will then analyze $\iota_1(\alpha)$ and $\iota_2(\alpha)$, respectively.

\subsection{Analysis of $\iota_1(\alpha)$}

For $\varepsilon(\alpha)$, we find that $
\varepsilon(\alpha)=\sum_{k=\alpha+1}^{\infty}\pi_k=\frac{1}{2}\Theta(\alpha)$. Thus, for the first term $\iota_1(\alpha)$ in Eq. \eqref{orierror}, we obtain 
\begin{equation}
	\begin{aligned}
		\lim_{\alpha\to+\infty}\iota_1(\alpha)
		&=\lim_{\alpha\to+\infty}\alpha\ln\left(1-\sum_{k=\alpha+1}^{\infty}\pi_k\right)\\
		&=\lim_{\alpha\to+\infty}-K\left(\sum_{k=K+1}^{\infty}\pi_k\right)  \\
		&=\lim_{\alpha\to +\infty}-\frac{1}{2}\alpha \Theta(\alpha).
	\end{aligned}\label{cond}
\end{equation}
This equation holds according to Eq. \eqref{cenchainprop}.  
From the above analysis, to ensure $\lim_{\alpha\to+\infty}\iota_1(\alpha)=0$, it is equivalent to let $\lim_{\alpha\to +\infty}\frac{1}{2}\alpha\Theta(\alpha)=0$.

In Lemma~\ref{lm4}, we shall prove that with $\Theta(\alpha)=2e^{-\theta\alpha+b\alpha^p}$, where $p\in(-\infty,1)$ and $b$ is a finite constant, $\lim_{\alpha\to +\infty}\frac{1}{2}\alpha\Theta(\alpha)=0$ holds.

\begin{lemma}\label{lm4}
	With $\Theta(\alpha)=2e^{-\theta\alpha+b\alpha^p}$, where $p\in(-\infty,1)$ and $b$ is a finite constant satisfying $-\theta k+bk^p\leq 0$ for $k=0,1,\cdots$, we have
	\begin{equation}
		\lim_{\alpha\to+\infty}\frac{1}{2}\alpha\Theta(\alpha)=0.	\end{equation}
\end{lemma}

\begin{IEEEproof}
	To ensure $\frac{1}{2}\Theta(\alpha)\leq 1$, $-\theta k+bk^p\leq 0$ for $k=0,1,\cdots$.
	With $p\in(-\infty,1)$ and a finite $b$ satisfying $-\theta k+bk^p\leq 0$ for $k=0,1,\cdots$, we obtain
	\begin{equation}
		\begin{aligned}
			\lim_{x\to+\infty}-\theta x+bx^p&=\lim_{x\to+\infty} x(-\theta+bx^{p-1})=-\infty.
		\end{aligned}
	\end{equation}
	
	Thus, we can then using L'Hospital's rule, which leads to
	\begin{equation}
		\begin{aligned}
			\lim_{\alpha\to+\infty}\frac{\alpha}{e^{\theta \alpha-b\alpha^p}}&=\lim_{\alpha\to+\infty}\frac{1}{(\theta-bp\alpha^{p-1})e^{\theta \alpha-b\alpha^p}}\\
			&=0.
		\end{aligned}
	\end{equation}
This equation holds since $\theta\alpha-b\alpha^p\to+\infty$.
\end{IEEEproof}

\begin{rem}
		\emph{Note that, to ensure the physical meaning of $\Theta(\alpha)$, there is another requirement on the value of $b$. Since $\Theta(\alpha)$ should decrease with increasing $\alpha$, $b$ should ensure $\Theta'(\alpha)\leq 0$. Therefore, to meet this requirement, $b$ can only be chosen from limited regions smaller than the mentioned one in Lemma~\ref{lm4}. Our analysis is still applicable in these limited regions.}
\end{rem}

Based on Lemma~\ref{lm4}, we have proved that with some specific requirements of $b$ and $p$, $\iota_1(\alpha)$ approaches zero as $\alpha\to+\infty$.

\subsection{Analysis of $\iota_2(\alpha)$}

For the second term $\iota_2(\alpha)$ in Eq. \eqref{orierror}, we find 
\begin{equation}\label{iota2def}
	\begin{aligned}
		\iota_2(\alpha)&=\ln\left(\frac{\prod_{k=0}^{\alpha-1} \sum_{i=k+1}^{\infty} \pi_i    }{ \prod_{k=0}^{\alpha-1}\left(\sum_{i=k+1}^{\infty}\pi_i-\sum_{i=\alpha+1}^{\infty}\pi_i\right)  }\right)\\
		&=\sum_{k=0}^{\alpha-1}\ln\left(1+\frac{\sum_{i=\alpha+1}^{\infty}\pi_i}{\sum_{i=k+1}^{\infty}\pi_i-\sum_{i=\alpha+1}^{\infty}\pi_i}\right).
	\end{aligned}	
\end{equation}

Since $\frac{x}{1+x}\leq \ln(1+x) \leq x $ holds for $x>-1$, we obtain 
\begin{equation}
	\sum_{k=0}^{\alpha-1}\frac{\sum_{i=\alpha+1}^{\infty}\pi_i}{ \sum_{i=k+1}^{\infty}\pi_i  }\leq \iota_2(\alpha)\leq 	\sum_{k=0}^{\alpha-1}\left(\frac{ \sum_{i=k+1}^{\infty}\pi_i }{\sum_{i=\alpha+1}^{\infty}\pi_i}-1\right)^{-1}. \label{upplowbd}
\end{equation}
For simplicity of notation, we define $\chi_1(\alpha)=\sum_{k=0}^{\alpha-1}\frac{\sum_{i=\alpha+1}^{\infty}\pi_i}{ \sum_{i=k+1}^{\infty}\pi_i  }$ and $\chi_2(\alpha)=\sum_{k=0}^{\alpha-1}\left(\frac{ \sum_{i=k+1}^{\infty}\pi_i }{\sum_{i=\alpha+1}^{\infty}\pi_i}-1\right)^{-1}$.
Thus, we can discuss $\iota_2(\alpha)$ from the perspective of its upper and lower bounds. Specifically, as $\alpha\to+\infty$, $\iota_2(\alpha)$ is a sum of infinite terms. If $\iota_2(\alpha)$ approaches a finite constant as $\alpha\to +\infty$, the last term must be infinitesimal.

We then consider the condition that  $\frac{1}{2}\Theta(\alpha)=e^{-\theta \alpha+b \alpha^p}$, where $p\in(-\infty,1)$ and $b$ is a finite constant satisfying $-\theta k+bk^p\leq 0$ for $k=0,1,\cdots$, and $\Theta'(x)\leq 0$ for $x\geq 0$. We summarize the result in Lemma~\ref{lm5}. The proof of Lemma~\ref{lm5} can be found in Appendix~\ref{lm5p}.

\begin{lemma}\label{lm5}
	With $\Theta(\alpha)=2e^{-\theta\alpha+b\alpha^p}$, where $p\in(-\infty,1)$ and $b$ is a finite constant satisfying $-\theta k+bk^p\leq 0$ for $k=0,1,\cdots$, and $\Theta'(x)\leq 0$ for $x\geq 0$, we have
	\begin{equation}
		0\leq\lim_{\alpha\to+\infty}\iota_2(\alpha)\leq \frac{1}{(e^{\theta}-1)(1-e^{-\theta })}.	
	\end{equation}
\end{lemma}

For $p=0$, i.e., $\Theta(\alpha)=ce^{-\theta \alpha}$, where $c=e^{b}$. The result in Lemma~\ref{lm5} can be improved. We summarize the improved results in Lemma~\ref{lm6}. The proof of Lemma~\ref{lm6} is provided in Appendix~\ref{lm6p}.

\begin{lemma}
	\label{lm6}
	With $p=0$, we have 
	\begin{equation}
		\frac{1}{e^{\theta}-1}\leq \lim_{\alpha\to+\infty}\iota_2(\alpha)\leq\frac{1}{(e^{\theta}-1)(1-e^{-\theta})}.	\end{equation}
\end{lemma}

Based on Lemmas~\ref{lm5} and~\ref{lm6}, we have obtained the bounds of $\iota_2(\alpha)$. Combing these results with Lemma~\ref{lm4}, we have proved that with $p\in(-\infty,0)\cup(0,1)$, 
\begin{equation}
	0\leq \lim_{\alpha\to+\infty}\vartheta(\alpha)\leq \frac{1}{(e^{\theta}-1)(1-e^{-\theta })}.
\end{equation}
With $p=0$, we have 
\begin{equation}
		\frac{1}{e^{\theta}-1}\leq \lim_{\alpha\to+\infty}\vartheta(\alpha)\leq \frac{1}{(e^{\theta}-1)(1-e^{-\theta })}.
\end{equation}

\section{Proof of Lemma~\ref{lm5}}\label{lm5p}
By substituting $\Theta(\alpha)=2e^{-\theta \alpha+b \alpha^p}$ into Eq. \eqref{upplowbd}, we obtain the upper and lower bounds. The lower bound is given by
\begin{equation}
	\chi_1(\alpha)=e^{-\theta \alpha} \sum_{k=0}^{\alpha-1}e^{\theta k}e^{b(\alpha^p-k^p)}.
\end{equation}

It is obvious that $\chi_1(\alpha)\geq 0$. Then, we analyze the upper bound $\chi_2(\alpha)$. The upper bound is given by 
\begin{equation}
	\chi_2(\alpha)=\sum_{k=0}^{\alpha-1}\frac{1}{e^{\theta(\alpha-k)}\cdot e^{b(k^p-\alpha^p)}-1}.
\end{equation}

For $x,y>1$, $\frac{1}{xy-1}<\frac{1}{(x-1)y}$. Since with the given $b$ satisfying the above mentioned conditions, $ e^{b(k^p-\alpha^p)}>1$. Thus, we obtain 
\begin{equation}
	\begin{aligned}
		\chi_2(\alpha)&\leq  \sum_{k=0}^{\alpha-1} \frac{1}{(e^{\theta(\alpha-k)}-1)e^{b(k^p-\alpha^p)}} \\
		&		\leq \sum_{k=0}^{\alpha-1}\frac{1}{(e^{\alpha-k}-1)e^{b\left((\alpha-1)^{p}-\alpha^p\right)}}. \label{uppldt2}
	\end{aligned}	
\end{equation}

Then, by combining Eq. \eqref{uppldt2} and Eq. \eqref{lowldt1}, we obtain
\begin{equation}
	\begin{aligned}
		\lim_{\alpha\to+\infty}\chi_2(\alpha)&\leq\lim_{\alpha\to+\infty}\frac{1}{e^{b\left((\alpha-1)^{p}-\alpha^p\right)}}\cdot\frac{1}{e^{\theta}-1} \sum_{i=1}^\alpha e^{-\theta(i-1)}\\
		&=\frac{1}{e^{\theta}-1}\lim_{\alpha\to+\infty}\frac{1}{e^{b\left((\alpha-1)^{p}-\alpha^p\right)}}\frac{1-e^{-\theta \alpha}}{1-e^{-\theta}}.
	\end{aligned}
\end{equation}
Note that
\begin{equation}
	\begin{aligned}
		\lim_{\alpha\to+\infty}(\alpha-1)^p-\alpha^p&=\lim_{\alpha\to+\infty}\alpha^p\left(\left(1-\frac{1}{\alpha}\right)^p-1\right)\\
		&=\lim_{\alpha\to+\infty}-p\alpha^{p-1}\\
		&=0.
	\end{aligned}
\end{equation}

Thus, we obtain that 
\begin{equation}
	\lim_{\alpha\to\infty}\chi_2(\alpha)\leq \frac{1}{(e^{\theta}-1)(1-e^{-\theta })}.
\end{equation}

The proof is completed.

\section{Proof of Lemma~\ref{lm6}}\label{lm6p}

By substituting $\Theta(\alpha)=c_2e^{-\theta \alpha}$ into Eq. \eqref{upplowbd}, we can derive the upper and lower bounds for $\iota_2(\alpha)$. Specifically, the lower bound is given by
\begin{equation}
	\begin{aligned}
		\chi_1(\alpha)&=\sum_{k=0}^{\alpha-1}\frac{e^{-\theta \alpha}}{ e^{-\theta k} }\\
		&=\frac{e^{-\theta \alpha}-1}{1-e^{\theta}}. \label{uppldt1}
	\end{aligned}	
\end{equation}
From Eq.~\eqref{uppldt1}, it follows that
\begin{equation}
	\lim_{\alpha\to +\infty}\chi_1(\alpha)=\frac{1}{e^{\theta}-1}.
\end{equation}

Next, the upper bound is expressed as
\begin{equation}
	\begin{aligned}
		\chi_2(\alpha)&=\sum_{k=0}^{\alpha-1}\frac{1}{e^{\theta(\alpha-k)}-1}\\
		&=\sum_{i=1}^{\alpha}\frac{1}{e^{\theta i }-1}.
	\end{aligned}
\end{equation}

For $i \geq 1$, it is straightforward to verify that $\frac{1}{e^{\theta i}-1} \leq \frac{1}{(e^{\theta}-1)e^{(i-1)\theta}}$. Thus, the upper bound satisfies
\begin{equation}\label{lowldt1}
	\begin{aligned}
		\lim_{\alpha\to+\infty}\chi_2(\alpha)&\leq \frac{1}{e^{\theta}-1}\lim_{\alpha\to+\infty} \sum_{i=1}^\alpha e^{-\theta(i-1)}\\
		&=\frac{1}{e^{\theta}-1}\lim_{\alpha\to+\infty}\frac{1-e^{-\theta \alpha}}{1-e^{-\theta}}\\
		&=\frac{1}{(e^{\theta}-1)(1-e^{-\theta})}. 
	\end{aligned} 
\end{equation}

Combining the above results, we conclude that
\begin{equation}
	\frac{1}{e^{\theta}-1} \leq \lim_{\alpha \to +\infty} \iota_2(\alpha) \leq \frac{1}{(e^{\theta}-1)(1-e^{-\theta})}.
\end{equation}

The proof is completed.

\section{Proof of Theorem~\ref{thm3}}

To prove Theorem~\ref{thm3}, we first select a threshold $d$, chosen sufficiently large such that $\Pr\{q[n] > d\}$ approaches $0$. We let $\eta_d=\Pr\{q[n]>d\}$. Combining Bayesian formula and Eq. \eqref{thm3e1}, $\varepsilon(\alpha)$ for $\alpha\geq d$ can be approximated as 
	\setcounter{equation}{80}
\begin{equation}\label{thm3p1}
	\varepsilon(\alpha)\approx \begin{cases}
		\eta_d\left(1+\frac{\tilde{\xi} (\alpha-d)}{\tilde{\sigma}}\right)^{-1/\tilde{\xi}}, \quad &\tilde{\xi}>0,\\
		\eta_de^{-(\alpha-d)/\tilde{\sigma}}, \quad &\tilde{\xi}=0.
	\end{cases}
\end{equation}

From Eq. \eqref{thm3p1}, we find that for $\tilde{\xi}=0$, the proof aligns with that of Theorem~\ref{thm2}. With $\tilde{\xi}=0$, the considered case is equivalent to setting $\theta=\frac{1}{\tilde{\sigma}}$, $p=0$, and $b=\ln \eta_d +\frac{d}{\tilde{\sigma}}$ in Theorem 2. Therefore, for $\tilde{\xi}=0$, we have 
\begin{equation}
 	\frac{1}{e^{\frac{1}{\tilde{\sigma}}}-1}\leq \lim_{\alpha\to+\infty}\vartheta(\alpha)\leq \frac{1}{(e^{\frac{1}{\tilde{\sigma}}}-1)(1-e^{-\frac{1}{\tilde{\sigma}} })}.
 \end{equation}

For $\tilde{\xi}> 0$, we have 
\begin{equation}\label{thm3p2}
	\begin{aligned}
		\lim_{\alpha\to+ \infty}\frac{1}{2}\alpha\Theta(\alpha)=&\lim_{\alpha\to+\infty}\eta_d\alpha\left(1+\frac{\tilde{\xi} (\alpha-d)}{\tilde{\sigma}}\right)^{-\frac{1}{\tilde{\xi}}}\\
	=&\lim_{\alpha\to+\infty}\frac{\eta_d\alpha}{\left(\alpha^{\tilde{\xi}}\left(\frac{\tilde{\sigma}-\tilde{\xi}d}{\tilde{\sigma}}\alpha^{-\tilde{\xi}}+\frac{\tilde{\xi}}{\tilde{\sigma}}\alpha^{1-\tilde{\xi}}\right)\right)^{\frac{1}{\tilde{\xi}}}}\\
	=&\lim_{\alpha\to+\infty}\frac{\eta_d}{\left(\left(1-\frac{\tilde{\xi} d}{\tilde{\sigma}}\right)\alpha^{-\tilde{\xi}}+\frac{\tilde{\xi}}{\tilde{\sigma}}\alpha^{1-\tilde{\xi}}\right)^{\frac{1}{\tilde{\xi}}}}.
	\end{aligned}
\end{equation}
From Eq. \eqref{thm3p2}, we deduce that for $\tilde{\xi}\in(0,1)$, $\lim_{\alpha\to+ \infty}\frac{1}{2}\alpha\Theta(\alpha)=0$. For $\tilde{\xi}=1$, $\lim_{\alpha\to+ \infty}\frac{1}{2}\alpha\Theta(\alpha)=\eta_d\tilde{\sigma}$, while for $\tilde{\xi}>1$, $\lim_{\alpha\to+ \infty}\frac{1}{2}\alpha\Theta(\alpha)=+\infty$.

Next, we will show that $\lim_{\alpha\to+\infty}\iota_2(\alpha)=+\infty$ for $\tilde{\xi}\in\left(\frac{1}{2},1\right]$. To this end, it suffices to prove that $\lim_{\alpha\to+\infty}\chi_1(\alpha)=+\infty$. We adopt the lower bound of $\chi_1(\alpha)$ as shown in Eq. \eqref{thm3p3}
	\setcounter{equation}{82}
\begin{equation}\label{thm3p3}
	\begin{aligned}
		\chi_1(\alpha)&=\sum_{k=0}^{\alpha-1}\frac{\sum_{i=\alpha+1}^{\infty}\pi_i}{ \sum_{i=k+1}^{\infty}\pi_i  } \\
		&=	\sum_{k=d}^{\alpha-1}\frac{\left(1+\frac{\tilde{\xi} (\alpha+1-d)}{\tilde{\sigma}}\right)^{-\frac{1}{\tilde{\xi}}}}{\left(1+\frac{\tilde{\xi} (k+1-d)}{\tilde{\sigma}}\right)^{-\frac{1}{\tilde{\xi}}}}\\
		&\geq\left(1+\frac{\tilde{\xi} (\alpha+1-d)}{\tilde{\sigma}}\right)^{-\frac{1}{\tilde{\xi}}} \sum_{k=d}^{\alpha-1}\left({1+\frac{\tilde{\xi} (k+1-d)}{\tilde{\sigma}}}\right)\\
			&= \!\!\left(\!1 \!+\!\frac{\tilde{\xi} (\alpha\!+\!1\!-\!d)}{\tilde{\sigma}}\right)^{\!\!\!-\frac{1}{\tilde{\xi}}}\!\!\!\left( \!\alpha-d \!+\!\frac{\tilde{\xi}}{\tilde{\sigma}}\frac{(\alpha\!-\!d)(\alpha\!+\!1\!-\!d)}{2} \right).
	\end{aligned}
\end{equation}

From Eq. \eqref{thm3p3}, it is evident that for $\tilde{\xi}\in\left(\frac{1}{2},1\right]$, $\lim_{\alpha\to+\infty} \chi_1(\alpha)=+\infty$. Consequently, we have proved that $\lim_{\alpha\to+\infty}\iota_2(\alpha)>\lim_{\alpha\to+\infty}\chi_1(\alpha)=+\infty$ with $\tilde{\xi}\in\left(\frac{1}{2},1\right]$. Combining this result with $\lim_{\alpha\to+\infty}\frac{1}{2}\alpha\Theta(\alpha)<+\infty$, we have proved that for $\tilde{\xi}\in\left(\frac{1}{2},1\right]$ the accumulated error is unbounded.

For $\tilde{\xi}\in(0,\frac{1}{2}]$, we derive:
\begin{equation}\label{eq84}
	\begin{aligned}
		\iota_2(\alpha)&=\sum_{k=0}^{\alpha-1}\ln\left(\frac{\sum_{i=k+1}^{\infty}\pi_i}{\sum_{i=k+1}^{\infty}\pi_i-\sum_{i=\alpha+1}^{\infty}\pi_i}\right)\\
		&=-\sum_{k=0}^{\alpha-1} \ln\left(1-\frac{\sum_{i=\alpha+1}^{\infty}\pi_i}{\sum_{i=k+1}^{\infty}\pi_i}\right)\\
	\end{aligned}
\end{equation}

Thus, we find that,  a necessary condition for $\lim_{\alpha \to +\infty} \iota_2(\alpha) < +\infty$ is:
	\setcounter{equation}{84}
\begin{equation}
	\lim_{\alpha\to+\infty}\ln\left(1-\frac{\sum_{i=\alpha+1}^{\infty}\pi_i}{\sum_{i=\alpha}^{\infty}\pi_i}\right)=0.
\end{equation} 
This implies:
\begin{equation}
	\lim_{\alpha\to+\infty}\frac{\sum_{i=\alpha+1}^{\infty}\pi_i}{\sum_{i=\alpha}^{\infty}\pi_i}=0.\label{0310}
\end{equation} 

However, for  $\tilde{\xi}\in(0,\frac{1}{2}]$, Eq. \eqref{0310} does not hold. Thus, for $\tilde{\xi}\in(0,\frac{1}{2}]$, $\lim_{\alpha\to+\infty}\iota_2(\alpha) = +\infty$. Combining this result with $\lim_{\alpha\to+\infty}\frac{1}{2}\alpha\Theta(\alpha)=0$, we we have proved that for $\tilde{\xi}\in(0,\frac{1}{2}]$ the accumulated error is unbounded.

Finally, we will prove that the accumulated error for $\tilde{\xi}>1$ is unbounded. For $\tilde{\xi}>1$, we have found that $\lim_{\alpha\to+\infty}\iota_1(\alpha)=-\infty$ according to Eqs. \eqref{cond} and \eqref{thm3p2}. Moreover, Eq. \eqref{0310} does not hold with $\tilde{\xi}=1$, which implies $\lim_{\alpha\to+\infty}\iota_2(\alpha)=+\infty$. Therefore, we have to analyze the limit of the sum of $\iota_1(\alpha)$ and $\iota_2(\alpha)$. In the following, we will show that the lower bound of $\iota_1(\alpha)+ \iota_2(\alpha)$, i.e., $\iota_1(\alpha)+\chi_1(\alpha)$, is infinite when $\alpha\to+\infty$.

For $\iota_1(\alpha)$ with $\tilde{\xi}>1$, we find that $\iota_1(\alpha)=-\eta_d\left(\frac{\tilde{\xi}}{\tilde{\sigma}}\right)^{-\frac{1}{\tilde{\xi}}}\alpha^{1-\frac{1}{\tilde{\xi}}}\left(1+o(1)\right)$ from Eq. \eqref{thm3p2}. For $\chi_1(\alpha)$ with $\tilde{\xi}>1$, based on Eq. \eqref{upplowbd} we have
\begin{equation}\label{x1}
	\begin{aligned}
			\chi_1(\alpha)
			&=
			\left(1+\frac{\tilde{\xi} (\alpha+1-d)}{\tilde{\sigma}}\right)^{-\frac{1}{\tilde{\xi}}}
			\sum_{k=d}^{\alpha-1}
			\left(1+\frac{\tilde{\xi} (k+1-d)}{\tilde{\sigma}}\right)^{\frac{1}{\tilde{\xi}}} \\
			&=
			\left(\frac{\tilde{\xi}}{\tilde{\sigma}}\alpha\right)^{-\frac{1}{\tilde{\xi}}}
			\left[1 + o(1)\right]
			\sum_{j=1}^{\alpha-d}
			\left(\frac{\tilde{\xi}}{\tilde{\sigma}}j\right)^{\frac{1}{\tilde{\xi}}}
			\left[1 + o(1)\right] \\
			&=
			\left(\frac{\tilde{\xi}}{\tilde{\sigma}}\right)^{-\frac{1}{\tilde{\xi}}}
			\alpha^{-\frac{1}{\tilde{\xi}}}
			\left[1 + o(1)\right]
			\cdot
			\frac{\tilde{\xi}^{1+\frac{1}{\tilde{\xi}}}}{(1+\tilde{\xi})\tilde{\sigma}^{\frac{1}{\tilde{\xi}}}}
			\alpha^{1+\frac{1}{\tilde{\xi}}}
			\left[1 + o(1)\right] \\
			&= \frac{\tilde{\xi}}{1+\tilde{\xi}} \alpha + o(\alpha).
	\end{aligned}
\end{equation}

Thus, we obtain that 
\begin{equation}\label{x2}
	\begin{aligned}
			&\lim_{\alpha\to+\infty}\iota_1(\alpha)+\chi_1(\alpha)\\
			=&\frac{\tilde{\xi}}{1+\tilde{\xi}} \alpha + o(\alpha)-\eta_d\left(\frac{\tilde{\xi}}{\tilde{\sigma}}\right)^{-\frac{1}{\tilde{\xi}}}\alpha^{1-\frac{1}{\tilde{\xi}}}\left(1+o(1)\right)\\
			=&\frac{\tilde{\xi}}{1+\tilde{\xi}} \alpha + o(\alpha)\\
			=&+\infty.
	\end{aligned}
\end{equation}

Thus, $\lim_{\alpha\to+\infty}\iota_1(\alpha)+\iota_2(\alpha)\geq \lim_{\alpha\to+\infty}\iota_1(\alpha)+\chi_1(\alpha)=+\infty$, which indicates the accumulated error for $\tilde{\xi}>1$ is unbounded. The proof is thus complete.

\begin{figure*}[b]
	\normalsize \hrulefill
	\setcounter{equation}{93}
	\begin{equation}\label{thm3p5}
		\begin{aligned}
			\chi_2(\alpha)
			&=\left(e^{\left(1+\frac{\xi}{\sigma}(\alpha+1-\mu)\right)^{-\frac{1}{\xi}}}-1\right)\sum_{k=0}^{\alpha-1}\frac{1}{1-e^{\left(1+\frac{\xi}{\sigma}(\alpha+1-\mu)\right)^{-\frac{1}{\xi}}-\left(1+\frac{\xi}{\sigma}(k+1-\mu)\right)^{-\frac{1}{\xi}}}}\\
			&\leq \left(e^{\left(1+\frac{\xi}{\sigma}(\alpha+1-\mu)\right)^{-\frac{1}{\xi}}}-1\right)\frac{\alpha}{1-e^{\left(1+\frac{\xi}{\sigma}(\alpha+1-\mu)\right)^{-\frac{1}{\xi}}-\left(1+\frac{\xi}{\sigma}(\alpha-\mu)\right)^{-\frac{1}{\xi}}}}\\
		\end{aligned}
	\end{equation}

\end{figure*}

\section{Proof of Theorem~\ref{thm4}}

According to the definition of the GEV distribution in Eq. \eqref{evt1}, for $\xi<0$, the GEV distribution is short-tailed, which is not the focus of our analysis. For $\xi>0$, the GEV distribution is heavy-tailed, whille $\xi=0$ corresponds to a light-tailed GEV distribution. Thus, in the following proof, we mainly focus on $\xi\geq 0$. For $\xi\geq 0$, we have
\setcounter{equation}{88} 
\begin{equation}
	\varepsilon(\alpha)\approx \begin{cases}
	   1- e^{-\left(1+\frac{\xi(z-\mu)}{\sigma}\right)^{-\frac{1}{\xi}}}, \quad &\xi>0,\\
        1-e^{-e^{-\frac{z-\mu}{\sigma}}}, \quad &\xi=0,
	\end{cases},
\end{equation}

For $\left\{q[n]\right\}$ satisfying Eq. \eqref{thm3e2} with $\xi>0$, we have 
\begin{equation}\label{t01}
	\begin{aligned}
		 \lim_{\alpha\to+\infty}\frac{1}{2}\alpha\Theta(\alpha)=&\lim_{\alpha\to+\infty}\alpha\sum_{k=\alpha+1}^{\infty}\pi_k\\
		=&\lim_{\alpha\to+\infty}\alpha\left(1-e^{-\left(1+\frac{\xi}{\sigma}(\alpha+1-\mu)\right)^{-\frac{1}{\xi}}}\right)\\
		=&\lim_{\alpha\to+\infty}\frac{\alpha}{\left(1+\frac{\xi}{\sigma}(\alpha+1-\mu)\right)^{\frac{1}{\xi}}}.
	\end{aligned}
\end{equation}

Thus, based on Eq. \eqref{t01}, we have 
\begin{equation}
		\lim_{\alpha\to+\infty}\frac{1}{2}\alpha\Theta(\alpha)=
		\begin{cases}
			0, \quad &\xi\in(0,1),\\
			\sigma, \quad &\xi=1,\\
			+\infty, \quad & \xi>1.
		\end{cases}
\end{equation}

For $\xi =0$, we have 
\begin{equation}
	\begin{aligned}
		\lim_{\alpha\to+\infty}\frac{1}{2}\alpha\Theta(\alpha)=&\lim_{\alpha\to+\infty}\alpha\left(1-e^{-e^{-\frac{\alpha+1-\mu}{\sigma}}}\right)\\
		=&\lim_{\alpha\to+\infty}\frac{\alpha}{e^{\frac{\alpha+1-\mu}{\sigma}}}\\
		=&0.
	\end{aligned}
\end{equation}

Next, we analyze $\iota_2(\alpha)$ for $\xi\in(0,1]$. We will show that $\chi_2(\alpha)$ is bounded, which indicates that $\iota_2(\alpha)$ is bounded. For $\chi_2(\alpha)$ with $\xi\in(0,1]$, we obtain
	\setcounter{equation}{92}
\begin{equation}\label{thm3p4}
	\begin{aligned}
		\chi_2(\alpha)&=\sum_{k=0}^{\alpha-1}\left(\frac{1-e^{-\left(1+\frac{\xi}{\sigma}(k+1-\mu)\right)^{-\frac{1}{\xi}}}}{1-e^{-\left(1+\frac{\xi}{\sigma}(\alpha+1-\mu)\right)^{-\frac{1}{\xi}}}}-1\right)^{-1}\\
		&=\sum_{k=0}^{\alpha-1}\frac{1-e^{-\left(1+\frac{\xi}{\sigma}(\alpha+1-\mu)\right)^{-\frac{1}{\xi}}}}{e^{-\left(1+\frac{\xi}{\sigma}(\alpha+1-\mu)\right)^{-\frac{1}{\xi}}}-e^{-\left(1+\frac{\xi}{\sigma}(k+1-\mu)\right)^{-\frac{1}{\xi}}}}.\\
	\end{aligned}
\end{equation}
By further manipulating Eq. \eqref{thm3p4}, we obtain Eq. \eqref{thm3p5}, which is located at the bottom of this page. By combining the results, we obtain
	\setcounter{equation}{94}
\begin{equation}\label{appxd1}
	\begin{aligned}
		\lim_{\alpha\to+\infty}\!\!\! \chi_2(\alpha)
		&\!\leq\! \lim_{\alpha\to+\infty}\!\! \frac{\alpha\left(1+\frac{\xi}{\sigma}(\alpha+1-\mu)\right)^{-\frac{1}{\xi}}}{\left(1+\frac{\xi}{\sigma}(\alpha-\mu)\right)^{\!\!-\frac{1}{\xi}}\!\!\!-\!\!\left(1+\frac{\xi}{\sigma}(\alpha+1-\mu)\right)^{\!\!-\frac{1}{\xi}}}\\
		&= \lim_{\alpha\to+\infty}\frac{\alpha}{\left(1-\frac{\frac{\xi}{\sigma}}{1+\frac{\xi}{\sigma}(\alpha+1-\mu)}\right)^{-\frac{1}{\xi}}-1}\\
		&=\lim_{\alpha\to+\infty}\frac{\sigma \alpha}{1+\frac{\xi}{\sigma}\left(\alpha+1-\mu\right)}\\
		&=\frac{\sigma^2}{\xi}.
	\end{aligned}
\end{equation}
Eq. \eqref{appxd1} holds since $\lim_{x\to 0}\frac{e^x-1}{x}=1$ and $\lim_{x\to0} \frac{(1-x)^{\beta}-1}{-\beta x}=1$. Based on Eq. \eqref{appxd1}, we conclude that $\lim_{\alpha\to+\infty}\iota_2(\alpha)\leq \frac{\sigma^2}{\xi}$ for $\xi\in(0,1]$. Therefore, we have proved that $\lim_{\alpha\to+\infty}\vartheta(\alpha)\leq  \frac{\sigma^2}{\xi} $ for $\xi\in(0,1)$ and $\lim_{\alpha\to+\infty}\vartheta(\alpha)\leq  \sigma^2+\sigma $ for $\xi=1$.

For $\xi=0$, according to Eq. \eqref{iota2def}, we have Eq. \eqref{appxde2}, which appears at the bottom of this page. For simplicity of notation, we define $\chi_3(\alpha)=\sum_{k=0}^{\alpha-1} \ln \left(1-e^{-e^{-\frac{k+1-\mu}{\sigma}}}\right)$ and $\chi_4(\alpha)=-\sum_{k=0}^{\alpha-1}\ln\left(e^{-e^{-\frac{\alpha+1-\mu}{\sigma}}}-e^{-e^{-\frac{k+1-\mu}{\sigma}}}\right)$.

\begin{figure*}[b]
	\normalsize \hrulefill
		\setcounter{equation}{95}
	\begin{equation}\label{appxde2}
		\begin{aligned}
			\iota_2(\alpha)
			&=\sum_{k=0}^{\alpha-1} \ln \left(1-e^{-e^{-\frac{k+1-\mu}{\sigma}}}\right) +\left[-\sum_{k=0}^{\alpha-1}\ln\left(e^{-e^{-\frac{\alpha+1-\mu}{\sigma}}}-e^{-e^{-\frac{k+1-\mu}{\sigma}}}\right)\right].
		\end{aligned}
	\end{equation}
	\setcounter{equation}{99}
	\begin{equation}\label{limitxi0}
		\begin{aligned}
			\lim_{\alpha\to+\infty}\iota_2(\alpha)
			&\leq \lim_{\alpha\to+\infty} e^{-\frac{1-\mu}{\sigma}}\cdot\frac{1-e^{-\frac{\alpha}{\sigma}}}{1-e^{-\frac{1}{\sigma}}}+ \alpha e^{-\frac{\alpha+1-\mu}{\sigma}}-\alpha -\alpha \ln\left(1-e^{e^{-\frac{\alpha+1-\mu}{\sigma}}-e^{-\frac{\alpha-\mu}{\sigma}}}\right)  \\
			&=\lim_{\alpha\to+\infty } e^{-\frac{1-\mu}{\sigma}}\cdot\frac{1-e^{-\frac{\alpha}{\sigma}}}{1-e^{-\frac{1}{\sigma}}}  +  \lim_{\alpha\to+\infty}\alpha e^{-\frac{\alpha+1-\mu}{\sigma}}-\alpha -\alpha \ln\left(1-e^{e^{-\frac{\alpha+1-\mu}{\sigma}}-e^{-\frac{\alpha-\mu}{\sigma}}}\right) \\
			&=\frac{e^{-\frac{1-\mu}{\sigma}}}{1-e^{-\frac{1}{\sigma}}}+\lim_{\alpha\to+\infty} \alpha e^{-\frac{\alpha+1-\mu}{\sigma}} +\alpha e^{e^{-\frac{\alpha+1-\mu}{\sigma}}-e^{-\frac{\alpha-\mu}{\sigma}}} -\alpha\\
			&=\frac{e^{-\frac{1-\mu}{\sigma}}}{1-e^{-\frac{1}{\sigma}}}+\lim_{\alpha\to+\infty}\alpha e^{-\frac{\alpha+1-\mu}{\sigma}} +\alpha \left(e^{-\frac{\alpha+1-\mu}{\sigma}}-e^{-\frac{\alpha-\mu}{\sigma}}\right)\\
			&=\frac{e^{-\frac{1-\mu}{\sigma}}}{1-e^{-\frac{1}{\sigma}}}.
		\end{aligned}
	\end{equation}
\end{figure*}

Then, we find that 
	\setcounter{equation}{96}
\begin{subequations}\label{defchi3}
	\begin{align}
		\chi_3(\alpha)&\leq - \sum_{k=0}^{\alpha-1} e^{-e^{-\frac{k+1-\mu}{\sigma}}} \label{defchi3a} \\
		& \leq - \sum_{k=0}^{\alpha-1} \left(1-e^{-\frac{k+1-\mu}{\sigma}}\right) \label{defchi3b} \\
		&= -\alpha+e^{-\frac{1-\mu}{\sigma}}\cdot\frac{1-e^{-\frac{\alpha}{\sigma}}}{1-e^{-\frac{1}{\sigma}}} \nonumber .
	\end{align}
\end{subequations}
Eq. \eqref{defchi3a} holds since $\ln(1-x)<-x$ for $0<x<1$. Moreover, Eq. \eqref{defchi3b} holds since $e^{-x}\geq 1-x $ for $0<x<1$. 

Additionally, we find that 
\begin{equation}\label{defchi4}
	\begin{aligned}
		\chi_4(\alpha)&=\alpha e^{-\frac{\alpha+1-\mu}{\sigma}}-\sum_{k=0}^{\alpha-1}\ln\left(1-e^{e^{-\frac{\alpha+1-\mu}{\sigma}}-e^{-\frac{k+1-\mu}{\sigma}}} \right)\\
		&\leq \alpha e^{-\frac{\alpha+1-\mu}{\sigma}} -\alpha \ln\left(1-e^{e^{-\frac{\alpha+1-\mu}{\sigma}}-e^{-\frac{\alpha-\mu}{\sigma}}}\right) .
	\end{aligned}
\end{equation} 

Besides, we have 
\begin{equation}\label{limitiota2}
	\begin{aligned}
		\lim_{\alpha\to+\infty}\iota_2(\alpha)&=\lim_{\alpha\to+\infty} \chi_3(\alpha)+\chi_4(\alpha) 
	\end{aligned}
\end{equation}
By substituting Eqs. \eqref{defchi3} and \eqref{defchi4} into Eq. \eqref{limitiota2}, we derive Eq. \eqref{limitxi0} as shown at the bottom of this page. 

Finally, we will prove that for $\xi>1$ the accumulated error is unbounded. With a similar method used for $\tilde{\xi}>1$ in Appendix D, we will show that the lower bound of $\iota_1(\alpha)+\iota_2(\alpha)$, i.e., $\iota_1(\alpha)+\chi_1(\alpha)$, is infinite when $\alpha\to+\infty$.

For $\iota_1(\alpha)$ with $\xi>1$, we find that $\iota_1(\alpha)=-\left(\frac{\xi}{\sigma}\right)^{-\frac{1}{\xi}}\alpha^{1-\frac{1}{\xi}}\left(1+o(1)\right)$ from Eq. \eqref{t01}. For $\chi_1(\alpha)$ with $\xi>1$, the following equation holds
\setcounter{equation}{100}
\begin{equation}
	\begin{aligned}
			\lim_{\alpha\to+\infty}\frac{1- e^{-\left(1+\frac{\xi(\alpha-\mu)}{\sigma}\right)^{-\frac{1}{\xi}}}}{\left(1+\frac{\xi(\alpha-\mu)}{\sigma}\right)^{-\frac{1}{\xi}}}=1.
	\end{aligned}
\end{equation}
Therefore, the analysis for $\chi_1(\alpha)$ is the same as the analysis for $\tilde{\xi}>1$ in Appendix B by setting $\tilde{\xi}=\xi$ and $d=\mu-1$. According to Eqs. \eqref{x1} and \eqref{x2}, we obtain
\begin{equation}
	\begin{aligned}
			\lim_{\alpha\to+\infty}\iota_1(\alpha)+\chi_1(\alpha)
			=\frac{\xi}{1+\xi} \alpha + o(\alpha)
			=+\infty.
	\end{aligned}
\end{equation}

Thus, $\lim_{\alpha\to+\infty}\iota_1(\alpha)+\iota_2(\alpha)\geq \lim_{\alpha\to+\infty}\iota_1(\alpha)+\chi_1(\alpha)=+\infty$, which indicates the accumulated error for $\xi>1$ is unbounded. The proof is thus complete.

\section{Proof of Theorem~\ref{thm5}}

To determine $Q_{\mathcal{A},\mathcal{A}}$, we need to derive $\Pr\left\{q[n+1]=j\kappa_q |q[n]=i\kappa_q \right\}$, where $i,j \in \{0,1, \cdots, \frac{\alpha}{\kappa_q}\}$. The conditions for $q[n+1]=j\kappa_q$ given $q[n]=i\kappa_q$ are given in Eq. \eqref{a1}, which is located at the top of the next page.

The transition probability can then be expressed as shown in Eq. \eqref{tp}, presented at the top of the next page.

For the case $s[n] \leq i\kappa_q$, substituting Eq. \eqref{a1} into Eq. \eqref{cap} provides the requirement for $P[n]$, as shown in Eq. \eqref{a2} at the top of the next page. By further substituting Eq. \eqref{lya} into Eq. \eqref{a2}, we derive the feasible interval for $|h[n]|^2$, denoted as $\mathcal{F}_{i,j}$. Note that the maximum queue length in the next time slot is $i\kappa_q + \lambda$ when the current queue length is $i\kappa_q$. This observation introduces the indicator in the first integral of Eq. \eqref{ijelements}.

For the case $s[n] > i\kappa_q$, the transmission rate exceeds the current queue length, which results in an empty buffer. Consequently, the queue length in the next time slot must be $\lambda$. This leads to the second integral in Eq. \eqref{ijelements}. The proof is complete.

\section{Proof of Theorem~\ref{thm6}}

From Eq. \eqref{lya}, we observe that $\left\lfloor\frac{C[n]}{A}\right\rfloor$ brings difficulty in the calculation of EC. To improve tractability,  we replace $\left\lfloor\frac{C[n]}{A}\right\rfloor$ with $\frac{C[n]}{A}$ in the subsequent derivations. This substitution leads to an approximation of $\theta_K$.

Consequently, Eq. \eqref{ecexpression} serves as both an approximation of ${\rm EC}_K(\theta_K)$ and an upper bound for ${\rm EC}_K(\theta_K)$, which is presented at the top of this page. By substituting Eq. \eqref{ecexpression} into the equation of $\theta_K$, we derive Eq. \eqref{thetaequ}. The proof is complete.

\begin{figure*}[t]
	\setcounter{equation}{102}
\begin{equation}\label{a1}
	\begin{cases}
		\kappa_q A \left(i-j+\frac{\lambda}{\kappa_q}\right) \!\! \leq C[n]<\!\! \kappa_q A \left(i-j+1+\frac{\lambda}{\kappa_q}\right), 	\!\! 	\!\!  &\text{if $s[n]\leq i\kappa_q$},\\
		j\kappa_q =\lambda, &\text{if $s[n]>i\kappa_q$}.		
	\end{cases}
\end{equation}
\begin{equation}\label{tp}
	\begin{aligned}
		\Pr\left\{q[n+1]=j\kappa_q \, | \, q[n]=i\kappa_q\right\}=& \mathbb{I}\left\{j\kappa_q=\lambda\right\}\Pr\left\{s[n]>i\right\}+ \\
		&\Pr\left\{s[n]\leq i, 	\kappa_q A \left(i-j+\frac{\lambda}{\kappa_q}\right) \leq C[n]< \kappa_q A \left(i-j+1+\frac{\lambda}{\kappa_q}\right)\right\}
	\end{aligned}
\end{equation}
\begin{equation}\label{a2}
	\frac{N_0B}{|h[n]|^2}\left(2^{\frac{\kappa_q A\left(i-j+\frac{\lambda}{\kappa_q}\right)}{BT}}-1\right)\leq P[n]<\min\left\{\frac{N_0B}{|h[n]|^2}\left(2^{\frac{\kappa_q A\left(i-j+1+\frac{\lambda}{\kappa_q}\right)}{BT}}-1\right),\frac{N_0B}{|h[n]|^2}\left(2^{\frac{\kappa_q A(i+1)}{BT}}-1\right)\right\}.
\end{equation}
	
	\begin{equation}\label{ecexpression}
		\begin{aligned}
			{\rm EC}_K(\theta_K)\approx & -\frac{1}{\theta_K}\ln\left[ \int_{0}^{+\infty}\left(1+\frac{x}{N_0B}\left(\frac{2BT}{VA}\omega \delta +\frac{2BT}{VA}\lambda -\frac{N_0B}{x}\right)^+\right)^{-\frac{BT}{A}\theta_K\log_2 e}f(x){\rm d}x\right] \\
			=&-\frac{1}{\theta_K}\ln\left[ \int_{0}^{\frac{N_0VA}{2T(\omega\delta+\lambda)}}f(x){\rm d} x+ \int_{\frac{N_0VA}{2T(\omega\delta+\lambda)}}^{+\infty}\left(\frac{2Tx}{N_0VA}\left(\omega \delta +\lambda\right) \right)^{-\frac{BT}{A}\theta_K\log_2 e}f(x){\rm d}x\right] \\
			=& -\frac{1}{\theta_K} \ln \left[ 1-\exp\left( -\frac{N_0VA}{2T\left(\omega \delta +\lambda\right)}  \right) +\left(  \frac{N_0VA}{2T(\omega \delta+\lambda)}  \right)^{\frac{BT}{A}\theta_K\log_2 e }\int_{  \frac{N_0VA}{2T(\omega\delta +\lambda)} }^{+\infty} x^{-\frac{BT}{A}\theta_K\log_2 e}e^{-x}{\rm d}x \right]\\
			=&-\frac{1}{\theta_K} \ln \left[ 1-\exp\left( -\frac{N_0VA}{2T\left(\omega \delta +\lambda\right)}  \right) +\left(  \frac{N_0VA}{2T(\omega \delta+\lambda)}  \right)^{\frac{BT}{A}\theta_K\log_2 e }\Gamma\left(1-\frac{BT}{A}\theta_K\log_2 e, \frac{N_0VA}{2T(\omega\delta+\lambda)}\right)\right] .
		\end{aligned}
	\end{equation}
		\hrulefill
\end{figure*}

\section{Proof of Corollary 1}
 Under the assumption that $K=\omega\delta$ is large, $\psi$ approaches $0$. Additionally, $\theta_K$ is a function of $K$. To explicitly illustrate the relationship between $\theta_K$ and $\omega$, we let $\Upsilon(\omega)=\theta_K$. For simplicity, we define $\Phi(\omega)=\frac{BT}{A}\Upsilon(\omega)\log_2e$.  Thus, we have 
	\setcounter{equation}{106}
\begin{subequations}\label{equi}
	\begin{align}
		&\lim_{\omega\to+\infty}\frac{1-e^{-\psi}+\psi^{\Phi(\omega)}\Gamma\left(1-\Phi(\omega), \psi\right)}{\psi+\frac{1}{\Phi(\omega)-1}\psi}\\
		=&\lim_{\psi\to0}\frac{1-e^{-\psi}}{\psi\frac{\Phi(\omega)}{\Phi(\omega)-1}}+\lim_{\psi\to0}\frac{\psi\cdot\frac{\Gamma\left(1-\Phi(\omega), \psi\right)}{\psi^{1-\Phi(\omega)}}}{\psi\frac{\Phi(\omega)}{\Phi(\omega)-1}} \\
		=&\left(1-\frac{1}{\Phi(\omega)}\right)\left(1+\lim_{\psi\to0}\frac{\Gamma\left(1-\Phi(\omega), \psi\right)}{\psi^{1-\Phi(\omega)}}\right) \nonumber \\
		=&\left(1-\frac{1}{\Phi(\omega)}\right)\left(1-\frac{1}{1-\Phi(\omega)}\right) \label{lim1}\\
		=&1.
	\end{align}
\end{subequations}
Eq. \eqref{lim1} holds because for $s<0$, the incomplete gamma function satisfies \cite{incgamma2}
\begin{equation}
	\lim_{x\to0}\frac{\Gamma(s,x)}{x^s}=-\frac{1}{s}.
\end{equation}

By substituting Eq. \eqref{equi} into Eq. \eqref{ecexpression}, we obtain 
\begin{equation}\label{equi2}
	\begin{aligned}
		&\lim_{\omega\to+\infty} {\rm EC}_K(\theta_K) \\
		=&-\lim_{\omega\to+\infty}\frac{1}{\Upsilon(\omega)}\ln\left[\psi\left(1+\frac{1}{\frac{BT}{A}\Upsilon(\omega)\log_2e-1}\right)\right].
	\end{aligned}
\end{equation}

Since ${\rm EC}_K(\theta_K)=\lambda$ and $\lambda$ is a finite constant, we obtain $\lim_{\omega\to+\infty}\Upsilon(\omega)=+\infty$. Thus, we can utilize the results shown in Eq. \eqref{equi2} to approximate ${\rm EC}_K(\theta_K)$. The approximation is given by 
\begin{equation}\label{ecapprox}
	\begin{aligned}
		{\rm EC}_K(\theta_K)&\approx-\frac{1}{\Upsilon(\omega)}\ln\left[\psi\left(1+\frac{1}{\frac{BT}{A}\Upsilon(\omega)\log_2e-1}\right)\right] \\
		&=-\frac{1}{\Upsilon(\omega)}\left[\ln\psi-\ln\left(1-\frac{A}{BT\Upsilon(\omega)\log_2 e}\right)\right].
	\end{aligned}
\end{equation}

By substituting Eqs. \eqref{ebe} and \eqref{ecapprox} into Eq. \eqref{thetaequ}, we have 
\begin{equation}
	\Upsilon(\omega)-\frac{1}{\lambda}\ln\left(1-\frac{A}{BT\Upsilon(\omega)\log_2e}\right)\approx -\frac{1}{\lambda}\ln \psi.
\end{equation}

The proof is complete.

\bibliographystyle{IEEEtran}
\bibliography{mybib}

\end{document}